\documentclass[10pt,onecolumn,draftcls]{packages/IEEEtran}
\pdfoutput=1

\usepackage{calrsfs,cite}

\usepackage{graphicx}
\usepackage{dsfont}
\usepackage{paralist}

\usepackage{color,calrsfs,amsthm,amsfonts,amssymb,amsbsy,amsmath}

   \newtheoremstyle{note}
     {10pt}
     {10pt}
     {\sf}
     {}
     {\bf}
     {:}
     {.5em}
     {}

\theoremstyle{note}

\usepackage{algorithm}
\usepackage{algorithmic}

\newtheorem{corollary}{Corollary}

\renewcommand\thepage{}

\usepackage{packages/myMathPack,packages/myPack,cite}

\usepackage{algorithm}
\usepackage{algorithmic}

\title{Mapping the Region of Entropic Vectors with Support Enumeration \& Information Geometry}

\author{Yunshu~Liu \quad John~MacLaren~Walsh \\
Dept. of ECE, Drexel University, Philadelphia, PA 19104, USA\\
yunshu.liu@drexel.edu \quad  jwalsh@coe.drexel.edu
\thanks{The authors thank the National Science Foundation for its support in part under CCF-1053702. Preliminary versions of this work were presented at the 2013 Information Theory Workshop and Allerton 2015.}
}

\cset{\Reals}{R} 
\scalar{\nv}{N} 
\gset{\ns}{N} 
\gset{\rvRangeSet}{X} 
\gset{\Kmc}{K} 
\gset{\Cir}{L} 
\gset{\matrixSet}{W} 

\rscalar{\rv}{X} 
\rvect{\rvv}{X}   
\scalar{\rank}{r} 
\vect{\rankVec}{r} 
\scalar{\entropy}{h} 
\scalar{\mutInf}{I} 
\scalar{\maxCard}{M} 
\scalar{\nAtoms}{k} 
\scalar{\nBlocks}{t} 
\scalar{\perm}{\pi}

\vect{\entVec}{h} 

\vect{\badRay}{f} 

\gset{\face}{F}

\family{\powSet}{P} 

\newcommand{\rvSupp}{\boldsymbol{\mathcal{X}}}

\newcommand{\setPartitionsOp}[1]{\Pi(#1)}
\newcommand{\setOneTo}[1]{\mathbb{N}_1^{#1}}

\newcommand{\partBlock}{B}
\gset{\setPartition}{B}
\newcommand{\subsetOfAllSetPartitions}{\xi}
\newcommand{\setOfSubsetsOfSetPartitions}{\boldsymbol{\Xi}}
\newcommand{\orbitsOp}[2]{#1 // #2}

\newcommand{\symGroup}{\mathbb{S}}

\newcommand{\meet}{\wedge}
\gset{\orbitTransversal}{T}

\begin{document}
\maketitle
\begin{abstract}
The region of entropic vectors is a convex cone that has been shown to be at the core of many fundamental limits for problems in multiterminal data compression, network coding, and multimedia transmission.  This cone has been shown to be non-polyhedral for four or more random variables, however its boundary remains unknown for four or more discrete random variables.  Methods for specifying probability distributions that are in faces and on the boundary of the convex cone are derived, then utilized to map optimized inner bounds to the unknown part of the entropy region.  The first method utilizes tools and algorithms from abstract algebra to efficiently determine those supports for the joint probability mass functions for four or more random variables that can, for some appropriate set of non-zero probabilities, yield entropic vectors in the gap between the best known inner and outer bounds.  These supports are utilized, together with numerical optimization over non-zero probabilities, to provide inner bounds to the unknown part of the entropy region.  Next, information geometry is utilized to parameterize and study the structure of probability distributions on these supports yielding entropic vectors in the faces of entropy and in the unknown part of the entropy region.
\end{abstract}
\vspace{-2mm}
\section{Introduction}
The region of entropic vectors is a convex cone that is known to be at the core of many yet undetermined fundamental limits for problems in data compression, network coding, and multimedia transmission.  This set has been shown to be non-polyhedral, but its boundaries remain unknown.  This paper studies methods of parameterizing probability distributions that yield entropic vectors in this unknown part of the entropy region.

In \S\ref{sec:rev}, after reviewing the definition of this set and its importance in applications, we discuss the best known outer and inner bounds for it, most of the latter of which are based on results for representable matroids and linear polymatroids.  These inner bounds, which, as we review, are based on linear constructions, are unable to parameterize the boundary of the region of entropic vectors because non-linear dependence structures are necessary to parameterize the unknown part of it.  

Bearing this in mind, in \S \ref{sec:atomgrown}, we provide a systematic and efficient method for searching for supports for joint probability mass functions which can, for appropriate choices of non-zero probabilities, yield entropic vectors in between the best known outer and inner bounds.  Key to this study is determining equivalence classes of supports, within which each of the contained supports are guaranteed, when ranging over all non-zero probabilities, to give the same sets of entropy vectors.  Only one representative support from each of these equivalence classes, the canonical representative, is necessary to be considered for the purposes of mapping entropies.  These equivalence classes are formalized via the language of group actions, and an efficient algorithm for listing canonical supports is provided.

With these canonical supports in hand, \S \ref{sec:experiments} sets about determining which of the supports for four random variables yield entropic vectors in the gap between the best known inner and outer bounds on the entropy region.  Via numerical optimization over the probabilities on these supports, tuned inner bounds to unknown part of the entropy region are also provided.

Seeking to gain a better understanding of the structure of probability distributions that would yield extremal entropic vectors in this unknown part of the entropy region, we shift in \S \ref{sec:infGeo} to studying the information geometric properties of probability distributions associated with entropic vectors that are extremal in the sense that they live in faces of the widely used outer bound.  Furthermore, those distributions on the smallest support that can yield entropic vectors in the unknown part of the entropy region are shown to have a special information geometric structure, which we also show disappears for larger supports having this property.  The paper concludes in \S \ref{sec:conculsions} with a number of interesting directions for further investigation.

\section{Bounding the Region of Entropic Vectors}\label{sec:rev}

Consider $N$ discrete random variables $\boldsymbol{X}=\boldsymbol{X}_{\mathcal{N}}=(X_1,\ldots,X_N),\ \mathcal{N}=\{1,\ldots,N\}$ with joint probability mass function $p_{\boldsymbol{X}}(\boldsymbol{x})$.  To every non-empty subset of these random variables $\boldsymbol{X}_{\mathcal{A}} := (X_n\ |\ n \in \mathcal{A}), \ \mathcal{A}\subseteq \mathcal{N}$, there is associated a Shannon entropy $H(\boldsymbol{X}_{\mathcal{A}})$ calculated from the marginal distribution $p_{\boldsymbol{X}_{\mathcal{A}}}(\boldsymbol{x}_{\mathcal{A}}) = \sum_{\boldsymbol{x}_{\mathcal{N}\setminus \mathcal{A}}} p_{\boldsymbol{X}_{\mathcal{N}}}(\boldsymbol{x})$ via
\begin{equation}
H(\boldsymbol{X}_{\mathcal{A}}) = \sum_{\boldsymbol{x}_{\mathcal{A}}} -p_{\boldsymbol{X}_{\mathcal{A}}}(\boldsymbol{x}_{\mathcal{A}}) \log_2 p_{\boldsymbol{X}_{\mathcal{A}}}(\boldsymbol{x}_{\mathcal{A}})
\end{equation}
One can stack these entropies of different non-empty subsets into a $2^N-1$ dimensional vector $\boldsymbol{h} = ( H(\boldsymbol{X}_{\mathcal{A}}) | \mathcal{A} \subseteq \mathcal{N})$, which can be clearly viewed as $\boldsymbol{h}(p_{\boldsymbol{X}})$, a function of the joint distribution $p_{\boldsymbol{X}}$.  A vector $\boldsymbol{h}_{?} \in \mathbb{R}^{2^N-1}$ is said to be \emph{entropic} if there is some joint distribution $p_{\boldsymbol{X}}$ such that $\boldsymbol{h}_{?} = \boldsymbol{h}(p_{\boldsymbol{X}})$.  The region of entropic vectors is then the image of the set $\mathcal{D} = \{ p_{\boldsymbol{X}} | p_{\boldsymbol{X}}(\boldsymbol{x}) \geq 0,\ \sum_{\boldsymbol{x}} p_{\boldsymbol{X}}(\boldsymbol{x}) =1 \}$ of valid joint probability mass functions under the function $\boldsymbol{h}(\cdot)$, and is denoted by
\begin{equation}
\Gamma^*_N = \boldsymbol{h}(\mathcal{D}) \subsetneq \mathbb{R}^{2^N-1}
\end{equation}
It is known that the closure of this set $\overline{\Gamma}^*_N$ is a convex cone \cite{Yeung_ITNETCODBOOK_08}, but surprisingly little else is known about this cone for $N\geq 4$.  Understanding the ``shape'' and boundaries of the set $\overline{\Gamma}^*_N$ is the subject of this paper. \\

The  fundamental importance of $\overline{\Gamma}^*_N$ lies in several contexts in signal processing, compression, network coding and information theory \cite{Yeung_ITNETCODBOOK_08}.
Firstly, the fundamental rate region limits of many data compression problems that otherwise remain unsolved can be directly expressed in terms of $\overline{\Gamma}^*_N$ and related entropy vector sets.  In particular, a class of problems studied by Zhang and Yeung under the name of distributed source coding for satellite communications \cite{Yeung_TIT_4_99,Yeung_ITNETCODBOOK_08} have rate regions directly expressible as projections of $\overline{\Gamma}^*_N$ after intersection with linear entropy equalities associated with certain Markov chain conditions.
Secondly, the multiple multicast capacity region for a lossless network under network coding can also be directly expressed in terms of $\overline{\Gamma}^*_N$, as was proved in \cite{Yan_ISIT_07} and \cite{Yeung_ITNETCODBOOK_08,Yan_TIT_09_12}.  The multi-source network coding (MSNC) problem generalizes the YZDSC problem by allowing for an arbitrary topology of intermediate encoding nodes between the source encoders and decoders. If one can determine the boundaries of $\overline{\Gamma}^*_N$ one can determine the capacity region of any network under network coding.  Furthermore, Chan and Grant \cite{Chan_ISIT_07,ChaGra2008a,ChaGra2008b} proved that every part of the (unknown) boundary of $\overline{\Gamma}^*_N$ has a network whose capacity region depends on it. Hence, they have established that the problem of determining the boundaries of $\overline{\Gamma}^*_N$ is equivalent to the problem of calculating the capacity region of every network under network coding. \\
\indent Even more generally, as \emph{all} achievable rate regions in information theory are expressed in terms of information measures between random variables obeying certain distribution constraints, they can be expressed as linear projections of entropy vectors associated with these \emph{constrained} random variables.  In the case of network coding and the distributed source coding problems introduced in the previous two sections, these constraints are embodied solely as entropy constraints and conditional entropies being zero, and hence can handle, after a great deal of proof, the constraints solely in entropy space and work with the region of unconstrained entropic vectors.  However, as one moves to more general multi-terminal information theory problems, particular distributions, either marginal or conditional, are typically specified, and in many cases there are also distortion/expectation of some function of a collection of the variables constraints.  Such constraints which can only be expressed in probability distribution space make the problem even harder, and a  modified region of entropic vectors $\overline{\Gamma}^*_N(\mathcal{C})$ will need to be employed incorporating distribution constraints $\mathcal{C}$.  However, given the wide array of multiterminal information theory whose rate regions can be directly expressed in terms of the simpler unconstrained region $\overline{\Gamma}^*_N$, it makes sense to attempt to bound and characterize it first.   Thus, the remaining sections of this paper will attempt to review what is known about bounding $\overline{\Gamma}^*_N$, as well as introduce some new relevant ideas from abstract algebra, combinatorics, and information geometry to understand these bounds, and enabling the creation of new even better bounds.

\subsection{Outer Bounds of $\overline{\Gamma}^*_N$}\label{ssec:boundsbest}
Viewed as a function $h_{\mathcal{A}} = H(\boldsymbol{X}_{\mathcal{A}})$ of the selected subset, with the convention that $h_{\emptyset}=0$, entropy is \emph{sub-modular} \cite{Yeung_ITNETCODBOOK_08,Zhang_TIT_07_98}, meaning that
\begin{equation}\label{eq:subMod}
h_{\mathcal{A}} + h_{\mathcal{B}} \geq h_{\mathcal{A}\cap \mathcal{B}} + h_{\mathcal{A}\cup \mathcal{B}} \quad \forall \mathcal{A},\mathcal{B} \subseteq \mathcal{N},
\end{equation}
and is also \emph{non-decreasing} and \emph{non-negative}, meaning that
\begin{equation}\label{eq:nonDec}
h_{\mathcal{K}} \geq h_{\mathcal{W}} \geq 0 \quad \forall \ \mathcal{W} \subseteq \mathcal{K} \subseteq \mathcal{N}.
\end{equation}
Viewed as requirements for arbitrary set functions (not necessarily entropy) the inequalities (\ref{eq:subMod}) and (\ref{eq:nonDec}) are known as the \emph{polymatroidal axioms} \cite{Yeung_ITNETCODBOOK_08,Zhang_TIT_07_98}, and a function obeying them is called the \emph{rank function} of a \emph{polymatroid}.  If a set function $\boldsymbol{f}$ that obeys the polymatroidal axioms (\ref{eq:subMod}) and (\ref{eq:nonDec}) additionally obeys
\begin{equation}\label{eq:matroidRankFunction}
f_{\mathcal{A}}\leq |\mathcal{A}|,\quad f_{\mathcal{A}} \in \mathbb{Z} \quad \forall \mathcal{A}\subseteq \mathcal{N}
\end{equation}
then it is the rank function of a \emph{matroid} on the ground set $\mathcal{N}$.

Since entropy must obey the polymatroidal axioms, the set of all rank functions of polymatroids forms an outer bound for the region of entropic vectors which is often denoted by
\begin{equation}
\label{gammaN}
\Gamma_N = \left\{ \boldsymbol{h} \left| \begin{array}{c} \boldsymbol{h} \in \mathbb{R}^{2^N-1} \\ h_{\mathcal{A}} + h_{\mathcal{B}} \geq h_{\mathcal{A}\cap \mathcal{B}} + h_{\mathcal{A}\cup \mathcal{B}}\ \forall \mathcal{A},\mathcal{B}\subseteq \mathcal{N} \\ h_{\mathcal{K}} \geq h_{\mathcal{W}} \geq 0 \quad \forall \ \mathcal{W}\subseteq \mathcal{K} \subseteq \mathcal{N}
 \end{array}  \right. \right\}
\end{equation}
In fact, any inequality in (\ref{gammaN}) can be expressed as a sum of the following two types of elemental inequalities\cite{Yeung_ITNETCODBOOK_08}
\begin{eqnarray}
\label{elemental_ineq}
 &&h_{\mathcal{N}} - h_{\mathcal{N} \backslash i} \geq 0 , i \in \mathcal{N} \\ \nonumber
 &&h_{i\mathcal{K}} + h_{j\mathcal{K}} - h_{\mathcal{K}} - h_{ij\mathcal{K}} \geq 0 , \text{for } i \neq j, \mathcal{K} \subset \mathcal{N} \backslash ij
\end{eqnarray}
The inequalities (\ref{elemental_ineq}) are the minimal, non-redudant, set of information inequalities for defining $\Gamma_N$. As we can see from the definition, $\Gamma_N$ is a polyhedron, and this polyhedral set is often known as the Shannon outer bound for $\overline{\Gamma}^*_N$ \cite{Yeung_ITNETCODBOOK_08,Zhang_TIT_07_98}. 

While in the low dimensional cases we have $\Gamma_2 = \Gamma^*_2$ and $\Gamma_3 = \overline{\Gamma}^*_3$, for $N\geq 4$, $\Gamma_N \neq \overline{\Gamma}^*_N$.
Zhang and Yeung first showed this in \cite{{Zhang_TIT_07_98}} by proving a new inequality among $4$ variables
{\small
\begin{equation} \label{eq:nonShanZY}
2 I(C;D) \leq I(A;B) + I(A;C,D) + 3 I(C;D|A) + I(C;D|B)
\end{equation}
}
\vspace{-5mm}
\\
which held for entropies and was not implied by the polymatroidal axioms, which they dubbed a \emph{non-Shannon type} inequality to distinguish it from inequalities implied by $\Gamma_N$.  For roughly the next decade a few authors produced other new non-Shannon inequalities \cite{MakMak2002,DouFre2006b}.
In 2007, Mat\'{u}\v{s} \cite{Matus_ISIT_2007} showed that $\overline{\Gamma}^*_N$ is not a polyhedron for $N\geq 4$.  The proof of this fact was carried out by constructing a sequence of non-Shannon inequalities, including

\vspace{-3mm}
{\small
\begin{eqnarray}
\label{matusIneq}
s[I(A;B|C) + I(A;B|D) + I(C;D) - I(A;B)]  \\ \nonumber
+ I(B;C|A) + \frac{s(s+1)}{2} [ I(A;C|B) + I(A;B|C)] \geq 0
\end{eqnarray}
}
\vspace{-4mm}

Notice (\ref{matusIneq}) is the same as Zhang-Yeung inequality (\ref{eq:nonShanZY}) when $s$ = 1. Additionally, the infinite sequence of inequalities was used with a curve constructed from a particular form of distributions to prove $\overline{\Gamma}^*_N$ is not a polyhedron.
Despite this result, even $\overline{\Gamma}^*_4$ is still not fully understood. Since then, many authors has been investigating the properties of $\overline{\Gamma}^{*}_N$ with the hope of ultimately fully characterizing the region \cite{Kaced13,Hassibi_ISIT_07,Xu_ISIT_08,Walsh_TIT_EntFunc_09sub,Csirmaz_13,Liu_ITW_13}.


\subsection{Inner Bounds of $\overline{\Gamma}^*_N$}\label{ssec:boundsbest}
Shifting from outer bounds to bounding from the inside, the most common way to generate inner bounds for the region of entropic vectors is to consider special families of distributions for which the entropy function is known to have certain properties.  \cite{Walsh_Allerton_09entFunc,Walsh_TIT_EntFunc_09sub,Walsh_Allerton_10,Li_Allerton_12} focus on calculating inner bounds based on special properties of binary random variables.  However, the most common way to generate inner bounds is based on inequalities for representable matroids  \cite{Hassibi_ITApres_10}, boolean polymatroids \cite{CPadro_07,Csirmaz_94} and subspace arrangements.\\

For the latter method, we first introduce some basics in linear polymatroids and the Ingeton inner bound.
Fix a $N' \geq N$, and partition the set $\{ 1,\ldots,N'\}$ into $N$ disjoint sets $\mathcal{Q}_1,\ldots,\mathcal{Q}_N$.  Let $\boldsymbol{U}$ be a length $m$ row vector whose elements are i.i.d. uniform over the finite field $GF(q)$, and let $\mathbf{T}$ be a particular $m\times N'$ deterministic matrix with elements in $GF(q)$.  Consider the $N'$ dimensional vector
\begin{equation*}\label{eq:vecValued}
\boldsymbol{Y} = \boldsymbol{U} \mathbf{T},\
\textrm{and define} \
\boldsymbol{X}_i = \boldsymbol{Y}_{\mathcal{Q}_i}, \ i \in\{1,\ldots, N\}.
\end{equation*}
The subset entropies of the random variables $\{\boldsymbol{X}_i\}$ obey
\begin{equation} \label{eq:entFuncRank}
H(\boldsymbol{X}_{\mathcal{A}}) = g(\mathcal{A}) \log_2(q) = \textrm{rank}\left( \left[ \mathbf{T}_{\mathcal{Q}_i} \left|  i \in \mathcal{A} \right. \right] \right) \log_2(q).
\end{equation}
A set function $\boldsymbol{g}(\cdot) $ created in such a manner is called a linear polymatriod or a subspace rank function.  It obeys the polymatroidal axioms, and is additionally proportional to an integer valued vector.  If the $\mathcal{Q}_i$ are all singletons and $N'=N$, then this set function is proportional (via $\log q$) to the rank function of a representable matroid \cite{Oxley_Matroid}.  Alternatively, when the sets $\mathcal{Q}_i$ are not singletons and $N'>N$, such a construction is clearly related to a representable matroid on a larger ground set\cite{Oxley_Matroid}.  Indeed, the subspace rank function vector is merely formed by taking some of the elements from the $2^{N'}-1$ representable matroid rank function vector associated with $\mathbf{T}$.  That is, set function vectors created via (\ref{eq:entFuncRank}) are ($ \log_2 q$ scaled) projections of rank function vectors of representable matroids.

Set functions capable of being represented in this manner for some $N',q$ and $\mathbf{T}$, are called subspace ranks in some contexts \cite{Hammer_JCSS_00,Matus_CPC_95,Dougherty_DiscreteMathSub_10}, while other papers effectively define a collection of vector random variables created in this manner a subspace arrangement \cite{Kinser_JCTSA_11}.

Define $\mathcal{I}_N$ to be the conic hull of all subspace ranks for $N$ subspaces. It is known that $\mathcal{I}_N$ is an inner bound for $\overline{\Gamma}^*_N$\cite{Hammer_JCSS_00}, which we name the subspace inner bound. So far $\mathcal{I}_N$ is only known for $N\leq 5$ \cite{Dougherty_DiscreteMathSub_10,Kinser_JCTSA_11}.  More specifically, $\mathcal{I}_2 = \overline{\Gamma}^*_2 = \Gamma_2$, $\mathcal{I}_3 = \overline{\Gamma}^*_3 = \Gamma_3$. As with most entropy vector sets, things start to get interesting at $N=4$ variables (subspaces).  For $N=4$, $\mathcal{I}_4$ is given by the Shannon type inequalities (i.e. the polymatroidal axioms) together with six additional inequalities known as \emph{Ingleton}'s inequality \cite{Hammer_JCSS_00,Ingleton_CMA_71,Matus_CPC_95} which states that for $N=4$ random variables
\begin{equation}\label{eq:ingleton}
Ingleton_{ij} \geq 0
\end{equation}
where 
\begin{eqnarray*}\label{ingleton_define}
Ingleton_{ij} &&= I(X_k;X_l|X_i) + I(X_k;X_l|X_j) \\ \nonumber
&&+ I(X_i;X_j)  - I(X_k;X_l) 
\end{eqnarray*}

\noindent Thus, $\mathcal{I}_4$ is usually called the Ingleton inner bound.

\subsection{Semimatroids and Faces of $\Gamma_N$ and $\Gamma_N^*$}
In \cite{Matus_CPC_95}, the concept of \emph{semimatroid} is introduced to help analyze the conditional independences among four random variables. In 
this section, we first review the definition of a \emph{semimatroid}; then some results in \cite{Matus_CPC_95} that are highly related to the structure of the region of entropic vectors on four variables will be presented; at last, we build the mapping between subset of extreme rays of $\Gamma_4$ and some particular $p$-$representable$ semimatroids, which we will use information geometry to analyze in section \S \ref{sec:infGeo}.\\

Let $\mathcal{N} = \{1,2,\cdots N\}$ and $\mathcal{S}$ be the family of all couples $(i,j | \mathcal{K})$, where $\mathcal{K} \subset \mathcal{N}$ and $ij$ is the union of two singletons $i$ and $j$ in $\mathcal{N}\setminus \mathcal{K}$. If we include the cases when $i = j$, there are, for example, 18 such couples for three variables, and 56 such couples for $N$ = 4. A relation $\mathcal{L} \subset \mathcal{S}$ is called \emph{probabilistically representable} or \emph{p-representable} if there exists a system of $N$ random variables $\mathbf{X} = \{X_i\}_{i \in \mathcal{N}}$ such that
\begin{eqnarray*}
&& \mathcal{L}= \{(i,j | \mathcal{K}) \in \mathcal{S}(N) | \text{$X_i$ is conditionally} \\
 && \quad \text{independent of $X_j$ given $X_\mathcal{K}$  i.e. $I(X_i ; X_j | X_{\mathcal{K}}) = 0$ $\}.$}
\end{eqnarray*}

\begin{definition} 
For $\boldsymbol{f} \in \Gamma_N$ we define $|[\boldsymbol{f}]|$ as
\begin{equation} \label{eq:semimatroid}
|[\boldsymbol{f}]| = \{(i,j | \mathcal{K}) \in \mathcal{S}(N) | f_{i\mathcal{K}} + f_{j\mathcal{K}} - f_{ij\mathcal{K}} -f_{\mathcal{K}} = 0 \}.
\end{equation}
A relation $\mathcal{L} \subset \mathcal{S}(N)$ is called a $semimatroid$ if and only if $\mathcal{L} = |[\boldsymbol{f}]|$ for some $\boldsymbol{f} \in \Gamma_N$, the Shannon outer bound for $N$ random variables. 
\end{definition}
We use $Semi(N)$ to denote the set of all semimatroids on $N$, and we say that semimatroid $\mathcal{L}$, $\mathcal{L} = |[\boldsymbol{f}]|$, \emph{arises from} polymatroid vector $\boldsymbol{f}$. 
The $p$-$representable$ semimatroids are just those semimatroids arising from an entropic vector $\boldsymbol{h}$. We use $P_{rep}(N)$ to denote the set of all $p$-$representable$ relations on $\mathcal{N}$. For $N \leq 3$, since $\Gamma_N = \overline{\Gamma}^{*}_N$, we have $P_{rep}(N) = Semi(N)$. However, $P_{rep}(4) \subsetneq Semi(4)$, that is to say, there are semimatroids on four variables that are not $p$-$representable$. The main theorem of \cite{Matus_CPC_99} lists all irreducible $p$-$representable$ semimatroids over four variables. There are 120 such semimatroids of 16 types, and every $p$-$representable$ semimatroid is at the intersection of some of these semimatroids. 
For $\mathcal{N} = \{1,2,3,4\}$, with $\mathcal{K} \subseteq \mathcal{N}$ and $0 \leq t \leq |\mathcal{N} \backslash \mathcal{K}|$, define $\boldsymbol{r}^{\mathcal{K}}_t$, $\boldsymbol{g}^{(2)}_i$ and $\boldsymbol{g}^{(3)}_i$ such that $\boldsymbol{r}^{\mathcal{K}}_t (\mathcal{W})$, $\boldsymbol{g}^{(2)}_i (\mathcal{W})$ and $\boldsymbol{g}^{(3)}_i (\mathcal{W})$ as follows: 
\[
\boldsymbol{r}^{\mathcal{K}}_t (\mathcal{W}) = \min \{ t, |\mathcal{W} \backslash \mathcal{K} | \} \ with \ \mathcal{W} \subseteq \mathcal{N}
\]
\[
\boldsymbol{g}^{(2)}_i (\mathcal{W}) =  \left\{
  \begin{array}{l l}
   2 & \quad \text{if $\mathcal{W} = i$}\\
   min \{ 2, |\mathcal{W}| \} & \quad \text{if $\mathcal{W} \neq i$}
   \end{array} \right.
\]
\[
\boldsymbol{g}^{(3)}_i (\mathcal{W}) =  \left\{
  \begin{array}{l l}
   |\mathcal{W}| & \quad \text{if $i \not \in \mathcal{W}$}\\
   min \{ 3, |\mathcal{W}|+1 \} & \quad \text{if $i \in \mathcal{W}$}
   \end{array} \right.
\]
Now we present the main theorem of \cite{Matus_CPC_99}
\begin{theorem}(Mat\'{u}\v{s})\cite{Matus_CPC_99}\label{theoremmatus}
There are 120 irreducible $p$-$representable$ semimatroids of sixteen types over four-element set $N$. Among which there are 36 ingleton semimatroids of 11 types:
$|[0]|$, $|[\boldsymbol{r}^{N-i}_1]|$ for $i \in \mathcal{N}$, $|[\boldsymbol{r}^{ij}_1]|$ for $i,j \in \mathcal{N}$ distinct, $|[\boldsymbol{r}^{i}_1]|$ for $i \in \mathcal{N}$, $|[\boldsymbol{r}_1]|$, $|[\boldsymbol{r}^{i}_2]|$ for $i \in \mathcal{N}$, $|[\boldsymbol{r}^{i \parallel j}_2]|$ for $i,j \in \mathcal{N}$ distinct, $|[\boldsymbol{r}_2]|$, $|[\boldsymbol{r}_3]|$, $|[\boldsymbol{g}^{(2)}_i]|$ for $i \in \mathcal{N}$, $|[\boldsymbol{g}^{(3)}_i]|$ for $i \in \mathcal{N}$.
There are also 84 non-Ingleton semimatroids of 5 types:\\
\begin{eqnarray*}
&&\mathcal{L}^{kl | \emptyset}_{ij} = \{(kl |i),(kl|j),(ij|\emptyset),(kl |ij)\}  \\
&& \quad  \quad \quad \cup \{(k|ij),(l|ij),(i|jkl),(j|ikl),(k|ijl),(l|ijk)\} \\
&&\mathcal{L}^{(ij | kl)}_{ij} = \{(ij|k),(ij|l),(kl|ij),(kl|i),(kl|j)\}\\
&&\mathcal{L}^{(ik|jl)}_{ij} = \{(kl|ij),(ij|k),(ik|l),(kl|j),(l|ij),(l|ijk)\} \\
&&\mathcal{L}^{ik|j}_{ij} = \{(ij|k),(ik|l),(kl|j),(i|jkl),(j|ikl)\}\\
&& \quad  \quad \quad \cup \{(k|ijl),(l|ijk)\} \\
&&\mathcal{L}^{jl | \emptyset}_{ij} = \{(kl |i),(jl|k),(ij|\emptyset),(kl|ij)\} \\
&& \quad  \quad \quad \cup \{(k|ij),(l|ij),(i|jkl),(j|ikl),(k|ijl),(l|ijk) \}
\end{eqnarray*}
\end{theorem}

The theorem not only solved $p$-\emph{representability} of semimatroids over four variables, but also answered the question of which faces of $\Gamma_4$ have interior points that are entropic, which is stated in the following corollary:
\begin{corollary} \label{thm:interiorpoints}
A relation $\mathcal{L}_0 \subseteq \mathcal{S}(N)$ is a $p$-$representable$ semimatroid if and only if
\begin{equation}
\label{prep_face}
\mathcal{F}_0 = \{ \ \boldsymbol{h} \in \Gamma_N \ | \ h_{i\mathcal{K}} + h_{j\mathcal{K}} - h_\mathcal{K} -h_{ij\mathcal{K}} = 0,\ \forall (i,j|\mathcal{K}) \in \mathcal{L} \}
\end{equation}
is a face of $\Gamma_N$ such that there exists $\boldsymbol{h}_0$, a point in the relative interior of $\mathcal{F}_*$, satisfying $\boldsymbol{h}_0 \in \Gamma^{*}_N$.
\end{corollary}
\noindent \textbf{Proof}: 
$\Rightarrow$ Suppose relation $\mathcal{L}_0$ is a $p$-$representable$ semimatroid. By definition (\ref{gammaN}) and (\ref{elemental_ineq}), 
\begin{equation}
\label{facetdefine}
\mathcal{F}_a = \{ \ \boldsymbol{h} \in \Gamma_N \ | \ h_{i\mathcal{K}} + h_{j\mathcal{K}} - h_\mathcal{K} -h_{ij\mathcal{K}} = 0  \}
\end{equation}
defines a facet of $\Gamma_N$ if $i\neq j$ or $i=j$ and $K=\mathcal{N}\setminus \{i,j\}$, and an intersection of such facets otherwise.  By definition a semimatroid $\mathcal{L}_0$ must be a union of $(i,j|\mathcal{K})$ couples such that there is a polymatroid $\boldsymbol{f} \in \Gamma_N$ which obeys exclusively these independence relations.  Thus $\mathcal{F}_0$ defined by any semimatroid must be face of $\Gamma_N$ because it is a exhaustive list of facets whose intersection forms this face.  Furthermore, since $\mathcal{L}_0$ is $p$-representibile, there exist some collection of random variables, generating an entropic vector in $\Gamma^*_n\cap \mathcal{F}_0$, that does not obey any additional conditional independence relations beyond $\mathcal{L}_0$.  This vector is thus in the relative interior of $\mathcal{F}_0$ (for being on a relative boundary of this face would require living in additional facets of $\Gamma_N$ and thus obeying additional conditional independence relations).  \\
$\Leftarrow$ Now suppose for a given $\mathcal{F}_0$, which is a face of $\Gamma_N$, we have $\boldsymbol{h}_0$, a relative interior point of $\mathcal{F}_0$ such that $\boldsymbol{h}_0 \in \Gamma^{*}_N$. Then the relation corresponding to the information equalities satisfied by $\boldsymbol{h}_0$ must be $p$-$representable$. \hspace*{\fill} $\Box$
 
\subsection{The gap between $\mathcal{I}_4$ and $\Gamma_4$}\label{ssec:gap}
For the structure of the gap between $\mathcal{I}_4$ and $\Gamma_4$, we know $\Gamma_4$ is generated by 28 elemental Shannon type information inequalities\cite{Yeung_ITNETCODBOOK_08}. As for $\mathcal{I}_4$, in addition to the the 28 Shannon type information inequalities, we also need six Ingleton's inequalities (\ref{eq:ingleton}), thus $\mathcal{I}_4 \subsetneq \Gamma_4$.
In \cite{Matus_CPC_95} it is stated that $\Gamma_4$ is the disjoint union of $\mathcal{I}_4$ and six cones $\{\boldsymbol{h} \in \Gamma_4 | Ingleton_{ij} < 0 \}$. The six cones $G^{ij}_{4}=\{\boldsymbol{h} \in \Gamma_4 | Ingleton_{ij} \leq 0 \}$ are symmetric due to the permutation of inequalities $Ingleton_{ij}$, so it is sufficient to study only one of the cones. Furthermore, \cite{Matus_CPC_95}  gave the extreme rays of $G^{ij}_{4}$ in Lemma \ref{lemmamatus} by using $\boldsymbol{r}^{\mathcal{K}}_t$, $\boldsymbol{g}^{(2)}_i$, $\boldsymbol{g}^{(3)}_i$ and the following functions $\boldsymbol{f}_{ij}$: \\

\noindent For $\mathcal{N} = \{1,2,3,4\}$, define $\boldsymbol{f}_{ij} (\mathcal{W})$ as follows:
\[
\boldsymbol{f}_{ij} (\mathcal{W}) =  \left\{
  \begin{array}{l l}
   3 & \quad \text{if $\mathcal{W} \in \{ik,jk,il,jl,kl\}$}\\
   min \{ 4, 2|\mathcal{W}| \} & \quad \text{otherwise}
   \end{array} \right.
\]

\begin{lemma}(Mat\'{u}\v{s})\cite{Matus_CPC_95}\label{lemmamatus}
The cone $G^{ij}_{4} = \{\boldsymbol{h} \in \Gamma_4 | Ingleton_{ij} \leq 0, i,j \in \mathcal{N} \text{ distinct}\}$ is the convex hull of 15 extreme rays. They are generated by the 15 linearly independent functions $\boldsymbol{f}_{ij}$, $\boldsymbol{r}^{ijk}_1$, $\boldsymbol{r}^{ijl}_1$, $\boldsymbol{r}^{ikl}_1$, $\boldsymbol{r}^{jkl}_1$, $\boldsymbol{r}^{\emptyset}_1$, $\boldsymbol{r}^{\emptyset}_3$, $\boldsymbol{r}^{i}_1$, $\boldsymbol{r}^{j}_1$, $\boldsymbol{r}^{ik}_1$, $\boldsymbol{r}^{jk}_1$, $\boldsymbol{r}^{il}_1$, $\boldsymbol{r}^{jl}_1$, $\boldsymbol{r}^{k}_2$, $\boldsymbol{r}^{l}_2$, where $kl = \mathcal{N}\backslash ij$.
\end{lemma}
\vspace{-2mm}
Note that among the 15 extreme rays of $G^{ij}_{4}$, 14 extreme rays $\boldsymbol{r}^{ijk}_1$, $\boldsymbol{r}^{ijl}_1$, $\boldsymbol{r}^{ikl}_1$, $\boldsymbol{r}^{jkl}_1$, $\boldsymbol{r}^{\emptyset}_1$, $\boldsymbol{r}^{\emptyset}_3$, $\boldsymbol{r}^{i}_1$, $\boldsymbol{r}^{j}_1$, $\boldsymbol{r}^{ik}_1$, $\boldsymbol{r}^{jk}_1$, $\boldsymbol{r}^{il}_1$, $\boldsymbol{r}^{jl}_1$, $\boldsymbol{r}^{k}_2$, $\boldsymbol{r}^{l}_2$ are also extreme rays of $\mathcal{I}_4$ and thus entropic, which leaves $\boldsymbol{f}_{ij}$ the only extreme ray in $G^{ij}_4$ that is not entropic\cite{Matus_CPC_95}.
It is easily verified that $\overline{\Gamma}^{*}_4$ is known as long as we know the structure of six cones $\overline{\Gamma}^{*}_4 \cap G^{ij}_{4}$. Due to symmetry, we only need to focus on one of the six cones $\overline{\Gamma}^{*}_4 \cap G^{34}_{4}$, thus we define $P^{34}_{4}=\overline{\Gamma}^{*}_4 \cap G^{34}_{4}$.  Thus, the remainder of the paper aims, in part, to study the properties of supports and probability distributions yielding entropic vectors in $P^{34}_{4}$, this gap between the best known inner and outer bounds for region of entropic vectors on four variables. 

Next, let's examine the relationship between subset of the extreme rays of $G^{34}_{4}$ and some particular $p$-$representable$ semimatroids. There are 15 extreme rays in $G^{34}_{4}$: $\boldsymbol{r}_1^{13}$, $\boldsymbol{r}_1^{23}$,  $\boldsymbol{r}_1^{123}$, $\boldsymbol{r}_1^{124}$, $\boldsymbol{r}_1^{134}$, $\boldsymbol{r}_1^{234}$, $\boldsymbol{r}_1^{\emptyset}$, $\boldsymbol{r}_1^{3}$, $\boldsymbol{r}_1^{4}$, $\boldsymbol{r}_1^{14}$, $\boldsymbol{r}_2^{1}$, $\boldsymbol{r}_1^{24}$, $\boldsymbol{r}_2^{2}$, $\boldsymbol{r}_3^{\emptyset}$, $\boldsymbol{f}_{34}$. We verified that none of the 56 $(i,j | \mathcal{K})$ couples is satisfied by all of the 15 extreme rays. If we remove $\boldsymbol{r}_1^{24}$, then $(1, 3 | 2)$ is the only relation satisfied by all the rest 14 extreme rays; if we remove both $\boldsymbol{r}_1^{24}$ and $\boldsymbol{r}_1^{14}$, then two relations $\{(1, 3 | 2) \ \& \ (2, 3 | 1)\}$ are satisfied by all the rest 13 extreme rays; at last if we remove $\boldsymbol{r}_1^{24}$, $\boldsymbol{r}_1^{14}$ and $\boldsymbol{r}_1^{3}$, then the remaining 12 extreme rays all satisfy the set of three relations $\{(1, 3 | 2) \ \& \ (2, 3 | 1) \ \& \ (1, 2 | 3)\}$.  From Theorem \ref{theoremmatus}, relations $\{(1,2|3)\}$, $\{(1,2|3) \ \& \ (2,3|1)\}$ and $\{(1,2|3) \ \& \ (1,3|2) \ \& \ (2,3|1)\}$ as semimatroids are all $p$-$representable$, we can say that the faces of $\Gamma_4$ generated by the corresponding subset of the extreme rays all have interior points that are $p$-$representable$. Furthermore, as we will see in \ref{ssec:infGeoSubMod}, the probability distributions associated with these faces of $\Gamma_4$ are well characterized in Information Geometry.

\section{Listing Canonical $\nAtoms$-atom Supports}\label{sec:atomgrown}
A first key question when studying the part of the entropy region associated with the gap between its best known inner and outer bounds described in the prior section is which supports for joint probability mass functions for the random variables can yield entropic vectors in this region.   In fact, results from Chan\cite{Chan_TIT_07_02} have shown that, with infinite computational power, to determine the whole entropy region, it would suffice to consider the conic hull of entropic vectors associated with only the \emph{quasi-uniform} probability distributions, which are completely specified via their support.  This question is further motivated by the observation that previous campaigns that have attempted to numerically map unknown parts of this region have empirically observed that the joint probability mass functions associated with the extremal entropic vectors produced, while not quasi-uniform, do have many of their probabilities zero \cite{DFZ_IneqsPaper,Csirmaz_13}. 

To begin our study in this arena, we must formally introduce the concept of a $k$-atom support and define the equivalence of two $k$-atom supports. Consider the probability distributions for a random vector $\boldsymbol{X}=(X_1,\ldots,X_N)$ taking values on the Cartesian product $\boldsymbol{\mathcal{X}}^{\times} = \mathcal{X}_1\times \mathcal{X}_2 \times \cdots \times \mathcal{X}_N$, where $\mathcal{X}_n$ is a finite set with values $i \in\{1,\ldots,|\mathcal{X}_n|\}$.  To a particular probability mass function $p_{\boldsymbol{X}}$ we can associate a length $\prod_{n=1}^N | \mathcal{X}_n| -1$ vector by listing the probabilities of all but one of the outcomes in $\boldsymbol{\mathcal{X}}^{\times}$ into a vector  
\begin{equation}
\label{eq:eta}
\boldsymbol{\eta} = \left[p_{\mathbf{X}}(i_1,\ldots,i_N)\left| \begin{array}{c} i_k \in\{1,2,\ldots,|\mathcal{X}_k|\}, \\ k\in\{1,\ldots,N\}, \\ \sum^{N}_{k=1} i_k \neq N. \end{array} \right. \right].
\end{equation}
Notice we are only listing outcomes such that $\sum_k i_k \neq N$, $p_{\mathbf{X}}(\bold{1}) = p_{\mathbf{X}}(i_1=1,\ldots,i_N=1)$ will be the only outcome that is left out. $\boldsymbol{\eta}$  in (\ref{eq:eta}) can be determined uniquely from the probability mass function $p_{\boldsymbol{X}}$, and owing to the fact that the probability mass function must sum to one, the omitted probability $p_{\mathbf{X}}(\bold{1})$ can be calculated, and hence the probability mass function can be determined from $\boldsymbol{\eta}$. 

Specifying the joint probability distribution via the vector $\boldsymbol{\eta}$ enables all outcomes to have nonzero probabilities, however, entropic vectors are often extremized by selecting some of the elements of $\boldsymbol{\mathcal{X}}^{\times}$ to have zero probability.  For this reason, rather than specifying some probabilities in their cartesian product to be zero, it is of interest to instead specify a support $\rvSupp\subset \boldsymbol{\mathcal{X}}^{\times}$, no longer a cartesian product, on which the probabilities will be non-zero.   Equivalently, if we take $|\rvSupp|=k$, we are considering only those probability spaces $(\Omega,\mathcal{F},\mathbb{P})$ with $|\Omega|=\nAtoms$ to define the random variables $\boldsymbol{X} : \Omega \rightarrow \boldsymbol{\mathcal{X}}^{\times} $ on.  A probability support $\rvSupp$ satisfying $|\rvSupp|=\nAtoms$ is called a $\nAtoms$-atom support, and a joint distribution created this way will be called a $\nAtoms$-atom distribution.  

Two $\nAtoms$-atom supports $\rvSupp,\rvSupp'$, $|\rvSupp| = |\rvSupp'| = k$, will be said to be \emph{equivalent}, for the purposes of tracing out the entropy region, if they yield the same set of entropic vectors, up to a permutation of the random variables.  In other words, $\rvSupp$ and $\rvSupp'$ are equivalent, if, for every probability mass function $p_{\rvv}:\rvSupp \rightarrow [0,1]$, there is another probability mass function $p_{\rvv'}: \rvSupp' \rightarrow [0,1]$ and a bijection $\pi : \ns \rightarrow \ns$ such that
\eq{\entropy_{\mathcal{A}}(p_{\rvv}) = \entropy_{\pi(\mathcal{A})}(p_{\rvv'}) \quad \forall \mathcal{A} \subseteq \ns . }
Take $N$ = 4 and $|\rvSupp|$ = 1 as a trivial example, since we only have one outcome/atom, it will have the probability of 1. In this way, different $1$-atom supports like $[(0,0,0,0)]$, $[(0,0,1,2)]$, $[(0,1,2,3)]$ and $[(2,5,7,9)]$ are equivalent because they all map to the same 15 dimensional entropic vector with all zero elements.

The goal of this section is to formalize this notion of equivalent supports with the use of tools from abstract algebra, then describe some methods for enumerating and listing one representative from each equivalence class of supports.

In this regard, the following definitions from the theory of group actions will be helpful. 
\begin{definition} 
Let $Z$ be a finite group acting on a finite set $\mathcal{V}$, a \emph{group action} is a mapping
\begin{equation*} \label{eq:groupact}
Z \times \mathcal{V} \rightarrow \mathcal{V}: (z, v) \mapsto zv
\end{equation*}
such that if $e$ is the identity in $Z$, $ev = v$ $\forall v \in \mathcal{V}$ and for any $z_1,z_2\in Z$, $z_2 z_1 v = (z_2 z_1) v$ for all $v\in V$.  For $v \in \mathcal{V}$, the \emph{orbit} of $v$ under $Z$ is defined as
\begin{equation*} \label{eq:orbits}
Z(v) = \{ zv \ | \ z \in Z \}
\end{equation*}
the \emph{stabilizer} subgroup of $v$ in $\mathcal{V}$ is defined as
\begin{equation*} \label{eq:stabilizer}
Z_v = \{ z \in Z \ | \ zv = v \}
\end{equation*}
Suppose there is some ordering of $\mathcal{V}$, and let $v$ be the element of $Z(v)$ that is least under this ordering, i.e. the \emph{canonical representative} of the orbit $Z(v)$.  For another $v^{\prime} \in Z(v)$, an element $z \in Z$ is called a \emph{transporter element} for $v^{\prime}$ if $zv^{\prime} = v$.
\end{definition}

\begin{definition}(orbit data structure \cite{BettenBook})
Let $Z$ be a group which acts on the finite set $\mathcal{V}$. The triple
\begin{equation*} \label{eq:obds}
orbit(Z,\mathcal{V}) = (\orbitTransversal, \sigma, \varphi)
\end{equation*}
is the \emph{orbit data structure} for $Z$ acting on $\mathcal{V}$ provided that
\begin{eqnarray*}
&&1. \orbitTransversal \text{ is a transversal of the $Z$-orbits on $\mathcal{V}$} \\
&&2. \sigma : \mathcal{V} \rightarrow L(Z) : v\mapsto Z_v \\
&&3. \varphi : \mathcal{V} \rightarrow Z : v \mapsto z \text{ with } zv \in \orbitTransversal
\end{eqnarray*}
Here, $L(Z)$ denotes the lattice of subgroups of $Z$, we call $\sigma$ the \emph{stabilizer map} and $\varphi$ the \emph{transporter map}.
\end{definition}
In next section, we will show that listing non-isomorphic distribution supports is equivalent to calculating the orbit data structure associated with the symmetric group acting a particular finite set.

\subsection{Non-isomorphic $\nAtoms$-atom supports via Snakes and Ladders}\label{ssec:snake}
\begin{figure}
\begin{center}
\includegraphics[width=.5\columnwidth]{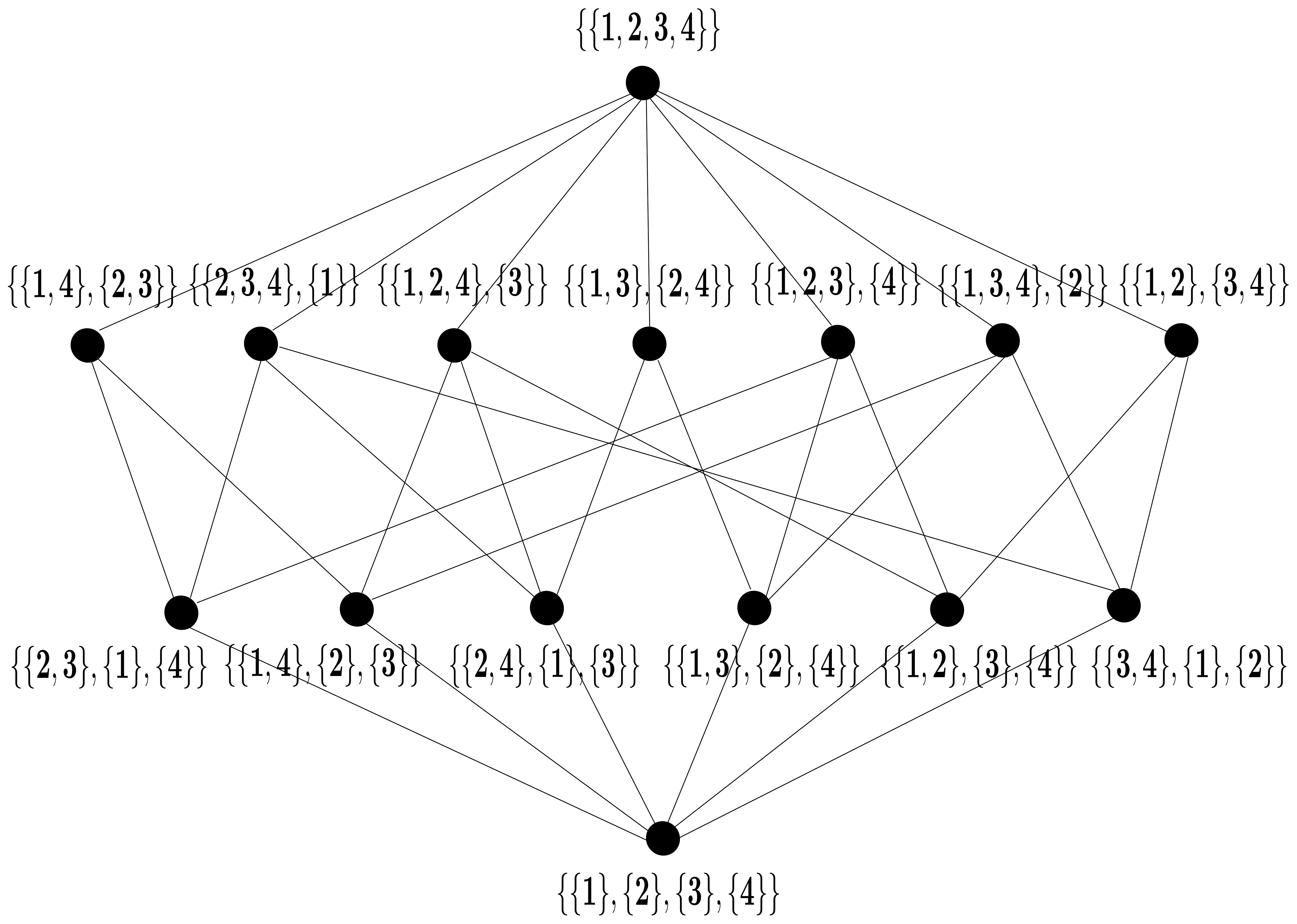}
\caption{Lattice of $\setPartitionsOp{\setOneTo{4}}$: the set of all set partitions for $k$ = 4}
\label{fg:lattice4}
\end{center}
\end{figure}

The key to list non-isomorphic distribution supports is to realize that a random variable on a probability space with $|\Omega| = \nAtoms$ can be viewed as a \emph{set partition} \cite{Knuth_ArtOfCP_11} of $\setOneTo{\nAtoms} = \{1,\ldots,\nAtoms\}$ for the purpose of calculating entropy.   A \emph{set partition} of $\setOneTo{\nAtoms}$ is a set $\setPartition = \{\partBlock_1,\ldots,\partBlock_{\nBlocks} \}$  consisting of $\nBlocks$ subsets $\partBlock_1,\ldots,\partBlock_{\nBlocks}$ of $\setOneTo{\nAtoms}$, called the \emph{blocks} of the partition, that are pairwise disjoint $\partBlock_i \cap \partBlock_j = \emptyset, \forall i\neq j$, and whose union is $\setOneTo{\nAtoms}$, so that $\setOneTo{\nAtoms} = \bigcup_{i=1}^{\nBlocks} \partBlock_i$.  Let $\setPartitionsOp{\setOneTo{\nAtoms}}$ denote the set of all set partitions of $\setOneTo{\nAtoms}$.  The cardinality of $\setPartitionsOp{\setOneTo{\nAtoms}}$ is commonly known as \emph{Bell numbers}.  For instance, there are 5 different set partitions for $k = 3$, that is $|\setPartitionsOp{\setOneTo{3}}|$ = 5 and 
\eq{\begin{array}{c} 
\setPartitionsOp{\setOneTo{3}} = \{\{\{1,2,3\}\},\{\{1,2\},\{3\}\},\{\{1,3\},\{2\}\}, \\
\{\{2,3\},\{1\}\}, \{\{1\},\{2\},\{3\}\} \},
\end{array} }
while for $k=4$, $|\setPartitionsOp{\setOneTo{4}}| = 15$ and $\setPartitionsOp{\setOneTo{4}}$ is the set
\eq{\begin{array}{c} 
\{ \{ \{1,2,3,4\} \}, \{ \{1,2,3\},\{4\} \},\{ \{1,2,4\},\{3\} \},\\ \{ \{1,3,4\},\{2\} \}, \{ \{2,3,4\},\{1\} \}, \{ \{1,2\},\{3,4\} \},\\ \{ \{1,3\},\{2,4\} \} , \{ \{1,4\}, \{2,3\}\}, 
\{ \{1,2\}, \{3\},\{4\} \},\\ \{ \{ 1,3\}, \{2\}, \{4\} \} ,\{ \{1,4\}, \{2\},\{3\} \}, \{ \{2,3\}, \{1\},\{4\} \},\\ \{ \{2,4\},\{1\},\{3\}\}, \{ \{3,4\}, \{1\},\{2\} \},
\{ \{1\},\{2\},\{3\},\{4\} \} \}.\end{array}
}
A set partition $\setPartition \in \setPartitionsOp{\setOneTo{\nAtoms}}$ is said to \emph{refine} a set partition $\setPartition' \in  \setPartitionsOp{\setOneTo{\nAtoms}}$ if all of the blocks in $\setPartition'$ can be written as the union of some blocks in $\setPartition$.  The \emph{meet} of two partitions $\setPartition,\setPartition'  \in \setPartitionsOp{\setOneTo{\nAtoms}}$, denoted by $\setPartition \meet \setPartition'$ is the partition of $\setOneTo{\nAtoms}$ formed by all of the non-empty intersections of a block from $\setPartition$ and a block from $\setPartition'$:
\eq{\setPartition \meet \setPartition' = \left\{ \partBlock_i \cap \partBlock_j' \left| \partBlock_i \in \setPartition, \partBlock_j' \in \setPartition', \partBlock_i \cap \partBlock_j' \neq \emptyset \right. \right\} }
Refinement and meet set up a partial order on $\setPartitionsOp{\setOneTo{\nAtoms}}$ which enable it to be identified as a \emph{lattice}, the lattice of $\setPartitionsOp{\setOneTo{4}}$ is shown in Figure \ref{fg:lattice4}

Let $\setOfSubsetsOfSetPartitions_{\nv}$ be the collection of all sets of $\nv$ set partitions of $\setOneTo{\nAtoms}$ whose meet is the finest partition (the set of singletons),
\begin{equation}
\label{eq:biggroup}
\setOfSubsetsOfSetPartitions_{\nv} := \left\{ \subsetOfAllSetPartitions \left| \subsetOfAllSetPartitions \subseteq \setPartitionsOp{\setOneTo{\nAtoms}},\ |\subsetOfAllSetPartitions| = \nv, \ \mathop{\bigwedge}_{\setPartition \in \subsetOfAllSetPartitions} \setPartition = \mathop{\bigcup}_{i=1}^{\nv} \{\{ i \}\} \right. \right\}.
\end{equation}

The symmetric group $\symGroup_{\nAtoms}$ induces a natural group action on a set partition $\setPartition \in \setPartitionsOp{\setOneTo{\nAtoms}}$,  $\setPartition = \{\partBlock_1,\ldots,\partBlock_{\nBlocks} \}$: representing an element $\perm \in \symGroup_{\nAtoms}$ as a permutation $\perm: \setOneTo{\nAtoms} \rightarrow \setOneTo{\nAtoms}$, we have
\begin{equation}
\label{eq:groupaction1}
 \perm ( \setPartition ) :=\left\{ \perm(\partBlock_1),\ldots,\perm(\partBlock_{\nBlocks}) \right\}.
\end{equation}
This action on set partitions induces, again in a natural manner, a group action of $\symGroup_{\nAtoms}$ on the set $\setOfSubsetsOfSetPartitions_{\nv}$ of subsets of $\nv$ partitions from $\setPartitionsOp{\setOneTo{\nAtoms}}$ whose meet is the singletons: $\perm \in \symGroup_{\nAtoms}$ acts on a set of partitions $\subsetOfAllSetPartitions \in \setOfSubsetsOfSetPartitions_{\nv}$, $\subsetOfAllSetPartitions = \{\setPartition_1,\ldots,\setPartition_{\nv}\}$ via
\begin{equation}
\label{eq:groupaction2}
\perm(\subsetOfAllSetPartitions) := \{ \perm(\setPartition_1),\ldots,\perm(\setPartition_{\nv}) \}.
\end{equation}
The group action (\ref{eq:groupaction1}) on set partition and group action (\ref{eq:groupaction2}) on sets of set partitions enable us to enumerate the non-isomorphic $\nAtoms$-atom supports by calculating the orbit data structure of symmetry group acting on a well defined set, the result is summarized in Theorem \ref{thm:enumEquiv}.
\thml{The problem of generating the list of all non-isomorphic $\nAtoms$-atom, $\nv$-variable supports, that is, selecting one representative from each equivalence class of isomorphic supports, is equivalent to obtaining a transversal of the orbits $\orbitsOp{\setOfSubsetsOfSetPartitions_{\nv}}{\symGroup_{\nAtoms}}$ of $\symGroup_{\nAtoms}$ acting on $\setOfSubsetsOfSetPartitions_{\nv}$, the set of all subsets of $\nv$ set partitions of the set $\setOneTo{\nAtoms}$ whose meets are the set of singletons $\{ \{1\},\{2\},\ldots,\{N\}\}$.}{thm:enumEquiv}

\textbf{Proof}: A random variable introduces a partition on the sample space based on its inverse image.  The joint distributions of several random variables is created from the meet of these partitions.  The joint entropy is insensitive to the labeling of elements in the sample space, as well as the labeling of the outcomes of the random variable, hence it is only the partitions and their meets that matter in determining the joint entropies.  Since an appropriate notion of isomorphism between supports also recognizes that it does not matter which random variable is labeled as the first random variable and so on, and there is no need to duplicate random variables when enumerating supports, rather than a $N$ tuple of partitions, the support of a collection of random variables is best thought of, then, as a set of such set-partitions. The requirement that the meet is the singletons follows from the fact that if it is not, there is a $\nAtoms'$ atom distribution with $\nAtoms'< \nAtoms$ whose atoms are the sets in the meet partition, which gives equivalent entropies, and hence such a situation is better labelled as a $\nAtoms'$-atom distribution.\hfill $\blacksquare$\\

Theorem \ref{thm:enumEquiv} sets up the theoretical framework for obtaining the list of non-isomorphic $\nAtoms$-atom, $\nv$-variable supports for the purpose of calculating entropic vectors: one must calculate the orbits data structure of the symmetric group ${\symGroup_{\nAtoms}}$ acting on $\setOfSubsetsOfSetPartitions_{\nv}$. One way to carry out this computation is to directly calculate the orbit data structure on $\setOfSubsetsOfSetPartitions_{\nv}$ using the default subroutine in GAP.  However, this approach quickly becomes intractable when $\nAtoms$ and $\nv$ are larger than four, as both CPU time and memory usage go beyond the resonable capacity of a single computer. Alternatively, one can use a recursive breadth-first search style algorithm named Leiterspiel or ``Snakes and Ladders''\cite{Schmalz_Bayreuth_92,BettenBook} to efficiently calculate the orbit data structure. 

Suppose we have a group $Z$ acting on a set $\mathcal{V}$, the algorithm Snakes and Ladders, see e.g. \cite{BettenBook} pp. 709--710, is an algorithm which enables one to compute orbit data structure of group $Z$ on the set $\mathcal{P}_i(\mathcal{V})$ of all subsets of the set $\mathcal{V}$ of cardinality $i$. For a given set $\mathcal{V}$, the algorithm first computes the orbit data structure on the set of subsets of size $i = 1$, then it recursively increase the subsets size $i$, where the computation to determine the orbit data structure for subsets of size $i$ is created from manipulations with the orbit data structure on subsets of size $i-1$.

To apply this problem to the non-isomorphic support enumeration problem, one selects the set $\mathcal{V}$ to be the set of all set partitions of the set $\setOneTo{\nAtoms}$, ordered lexicographically, and the group $Z$ to be the symmetric group $\symGroup_{\nAtoms}$.  Once the step in Snakes and Ladders associated with subsets (of set partitions) of size $N$ is reached, each element of the transversal is checked to determine if the meet of its partitions is the set of singletons, and the resulting sets of canonical $N$ set partitions yield the non-isomorphic supports.  

The snakes and ladders algorithm was applied to enumerate non-isomorphic supports in this manner, and the resulting numbers of non-isomorphic supports obtained are displayed in Table \ref{atomgrown}. As shown, for $N$ = 4 variables, only one support is needed for calculating entropic vectors of $3$-atom distribution, however, there are 75 non-isomorphic supports for $k$ = 4, and the number of non-isomorphic supports grow rapidly in the number of atoms $k$. 

In the next section, we will utilize these non-isomorphic $k$-atom supports together with numerical optimization to obtain inner bounds for entropy.

\begin{table}
\centering
\begin{tabular}{c|ccccc}
$\nv \backslash \nAtoms$ &  3 & 4 & 5 & 6 & 7 \\
\hline
2                        &  2  & 8   & 18     &  48 & 112 \\
3                        &  2  & 31  & 256    & 2437 & 25148 \\
4                        &  1  & 75  & 2665  & 105726 & 5107735 \\
5                        &  0  & 132 &  22422  &  3903832  &   \\
6                        &  0  & 187 &  161118   &  &   \\
\end{tabular}
\caption{\# of non-isomorphic $\nAtoms$-atom, $\nv$-variable supports.} \label{atomgrown}
\end{table}

\section{Mapping the Entropy Region with $k$-Atom Distributions}\label{sec:experiments}
With the list of canonical $\nAtoms$ supports obtained through the method described in the previous section in hand, the matter turns to how to exploit them to better numerically map out the unknown parts of the entropy region.  In this section, we study this problem from two angles, the first, in \S \ref{ssec:ingleVio}, aims to solely focus on optimizing the Ingleton score, while the second, in \S \ref{ssec:inner} describes a process for obtaining numerically optimized inner bounds to the entropy region.
\subsection{Maximal Ingleton Violation and the Four Atom Conjecture}\label{ssec:ingleVio}
Given that the unknown part of $\bar{\Gamma}^*_4$ is associated with violating the Ingleton inequality, substantial research effort has been exerted towards determining distributions on $N$ = 4 random variables that violate the Ingleton inequality (\ref{eq:ingleton}). Dougherty, Freiling, and Zeger\cite{DFZ_IneqsPaper} defined a normalized function called the \emph{Ingleton score} to measure the degree of Ingleton violation for 4 random variables, and they also make the \emph{Four-Atom Conjecture} which states that the Ingleton score of 4 random variables can not be lower than $-0.08937$. After the Four-Atom conjecture was proposed, Ingleton violation was studied extensively with finite groups\cite{Mao_ISIT10,Boston_Allerton_12,Pirita_TIT_07_14}, then in \cite{Csirmaz_13}, the conjecture was refuted by transforming a distribution obtaining Ingleton score of $-0.078277$ through a operation which preserves the property of almost entropic to a vector with Ingleton score $-0.09243$. In this section, we study the number of $\nAtoms$ atom supports that can, for some probability distribution, violate Ingleton, as well as the Ingleton scores they can attain.

For a particular $\nAtoms$ atom support for $\nv$ variables, the Ingleton score can be numerically optimized via fine grid search and numerical gradient optimization.  Doing so for each four variable support with $7$ or fewer atoms yielded the results in Table \ref{atomviolate}, which shows that only a small fraction of the canonical supports can violate Ingleton.

%
%
%

\begin{table}
\centering
\begin{tabular}{c|ccccc}
number of atoms $\nAtoms$ &  3 & 4 & 5 & 6 & 7\\
\hline
\text{all supports}                        &  1  & 75  & 2665  & 105726 & 5107735\\
\text{Ingleton violating}    &  0  &  1  &  29  &  1255 & 60996\\
\end{tabular}
\caption{\# of non-isomorphic $\nAtoms$-atom, $4$-variable supports that can violate the Ingleton inequality.} \label{atomviolate}
\end{table}

Among all the 75 non-isomorphic $4$-atom distribution supports, only one can be assigned a series of probabilities to violate Ingleton, that is the $4$-atom support (\ref{eq:4atoms}), 
\begin{eqnarray}
\label{eq:4atoms}
\begin{bmatrix}
(0,0,0,0) \\
(0,1,1,0) \\
(1,0,1,0) \\
(1,1,1,1)
\end{bmatrix},
\end{eqnarray}
which is the support achieving Ingleton score $-.08937$ associated with the Four atom conjecture.

Each row in (\ref{eq:4atoms}) is a vector in $\boldsymbol{\mathcal{X}}$ and corresponds to one outcome/atom for the collection of random variables, so the number of columns in (\ref{eq:4atoms}) is the same as the number of random variables, in this case $4$.   Among the 29 $5$-atom supports that can violate Ingleton, 28 of them obtain the same minimal Ingleton score of $-0.08937$, with one atom's probability shrinking to zero.  These $28$ thus all shrink to the $4$-atom support (\ref{eq:4atoms}), achieving the same Ingleton score.  The remaining one support, 
\begin{eqnarray}
\label{eq:5atoms}
\begin{bmatrix}
(0,0,0,0) \\ 
(0,0,1,1) \\
(0,1,1,0) \\
(1,0,1,0) \\
(1,1,1,0)
\end{bmatrix}
\end{eqnarray}
only achieves a minimal Ingleton score of $-0.02423$. For the 1255 $6$-atom supports, 58 of them get a minimal Ingleton score strictly less than $-0.08937$, while the remainder of the supports yield minimal scores of $-0.08937$.  Experiments with the $7$-atom supports have shown that, in keeping with the findings of the four atom conjecture and the attempts to refute it, no $\nAtoms$-atom support with $\nAtoms \leq 7$ is capable of directly beat the four atom distributions score.  These exhaustive results substantiate findings from other researchers that suggest that if it is indeed possible to give a probability distribution which directly (without any almost entropic preserving transformation on the entropic vectors as utilized in \cite{Csirmaz_13}) violates the four atom conjecture, at least a large support will be required.

\subsection{Optimizing Inner Bounds to Entropy from $k$-Atom Distributions}\label{ssec:inner}
For the purpose of generating better inner bounds for the region of entropic vectors, minimizing only the Ingleton score is far from enough, since it is only optimizing the distribution to a cost function of certain hyperplane defined by the Ingleton inequality. Bearing this in mind, one can define cost functions different from the Ingleton score, but still yielding optimized points that are in the unknown part of the entropy region associated with violating Ingleton.   We will first describe a simple procedure to randomly generate such cost functions, and then their numerical optimization over each of the Ingleton violating 4,5, and 6 atom supports.  The resulting entropic vectors are then collected to generate inner bounds to $\bar{\Gamma}^*_4$ based on distributions with 4,5, and 6 atom supports. 

Lemma \ref{lemmamatus} defined the 15 extreme rays of the pyramid $G^{ij}_{4}$, and, without loss of generality, it suffices to consider $G^{34}_{4}$. Among these 15 rays, the 14 extreme rays that lie on the hyperplane of $Ingleton_{34}$ = 0 are $\boldsymbol{r}^{134}_1$, $\boldsymbol{r}^{234}_1$, $\boldsymbol{r}^{123}_1$, $\boldsymbol{r}^{124}_1$, $\boldsymbol{r}^{\emptyset}_1$, $\boldsymbol{r}^{\emptyset}_3$, $\boldsymbol{r}^{3}_1$, $\boldsymbol{r}^{4}_1$, $\boldsymbol{r}^{13}_1$, $\boldsymbol{r}^{14}_1$, $\boldsymbol{r}^{23}_1$, $\boldsymbol{r}^{24}_1$, $\boldsymbol{r}^{1}_2$, $\boldsymbol{r}^{2}_2$, while the only extreme ray in $G^{34}_{4}$ that is not entropic is $\boldsymbol{f}_{34}$. For generating cost functions, among several options, we found the following one gives us the best inner bound. First a random vector $\Lambda = \{\lambda_1, \lambda_2, \cdots \lambda_{14}\}$ of 14 dimension is generated, where $\lambda_i$ takes value from 0 to a large positive value. Then for each of the 14 extreme rays on the hyperplane of  $Ingleton_{34}$ = 0, one new ray is generated through the following equation
\begin{equation}
\label{ }
\boldsymbol{r}^{new}_{i} = \frac{\boldsymbol{r}^{base}_{i} + \lambda_i \boldsymbol{f}_{34}}{1 + \lambda_i}
\end{equation}
where $\boldsymbol{r}^{base}_{i} \in  \{$$\boldsymbol{r}^{134}_1$, $\boldsymbol{r}^{234}_1$, $\boldsymbol{r}^{123}_1$, $\boldsymbol{r}^{124}_1$, $\boldsymbol{r}^{\emptyset}_1$, $\boldsymbol{r}^{\emptyset}_3$, $\boldsymbol{r}^{3}_1$, $\boldsymbol{r}^{4}_1$, $\boldsymbol{r}^{13}_1$, $\boldsymbol{r}^{14}_1$, $\boldsymbol{r}^{23}_1$, $\boldsymbol{r}^{24}_1$, $\boldsymbol{r}^{1}_2$, $\boldsymbol{r}^{2}_2\}$. After obtaining the 14 new rays $\mathbf{r^{new}} = \{\boldsymbol{r}^{new}_1, \cdots, \boldsymbol{r}^{new}_{14}\}$, the hyperplane defined by these new rays, which in turn defines a new cost function, can be easily calculated. Notice if we let $\lambda_i$ = 0 for $i = 1,2, \cdots 14$, we will get the hyperplane of $Ingleton_{34}$ = 0.

Computer experiments were run to generate more than 1000 cost functions in this manner by random selection of the $\lambda_i$s. For each of these cost functions, numerical optimization of the distribution for each Ingleton violating $k$-atom support was performed, and a $k$-atom inner bound was generated by taking the convex hull of the entropic vectors corresponding to the optimized distribution. 

The progress of characterizing $\Gamma^{*}_4$ while performing these experiments was in part estimated by the volume of the inner bound as compared to the total volume of the pyramid $G^{34}_{4}$, as summarized in Table \ref{tb:volume}. For the purpose of comparison, we also list there the calculated volume ratio of the best outer bound in \cite{DFZ_IneqsPaper}. Note the volume of the $k$-atom inner bound obtained through the process describe above is only a estimated value and a lower bound to the true volume fraction, because only a finite number of cost functions and a finite number of entropic vectors were generated through the random cost function generation process.  In principle one can generate as many entropic vectors as one wants through this process by growing the number of random cost functions selected, however calculating volume for many extreme points in high dimension can become computationally intractable.  A key observation from the process is that while growing the support helps, from a volume standpoint the improvement after four atoms is somewhat small.

\begin{table}
\centering
\begin{tabular}{c|ccccc}
inner and outer bounds &  percent of pyramid  \\
\hline
\text{Shannon}                        &  100 \\
\text{Outer bound from \cite{Dougherty_DiscreteMathSub_10}}    &  96.5 \\
\text{4,5,6 atoms inner bound}   & 57.8 \\
\text{4,5 atoms inner bound}   & 57.1\\
\text{4 atoms inner bound}   &  55.9\\
\text{4 atom conjecture point only}  &  43.5\\
\text{3 atoms inner bound}   &  0\\
\end{tabular}
\caption{The volume increase within pyramid $G^{34}_{4}$ as more atoms are included} \label{tb:volume}
\end{table}

However, volume is just one metric for an inner bound, which can also be hard to visualize in high dimensions.  For this reason, we would like to visualize the $k$-atom inner bound in lower dimension.  In this regard, a certain 3 dimensional subset of it selected in \cite{Csirmaz_13} will be utilized.  In order to perform the transformation, each numerically obtained 15 dimensional vector $\boldsymbol{h} \in  G^{34}_{4}$ is first transformed into its \emph{tight} component by subtracting its \emph{modular} component which is defined by
\begin{equation*}
\label{ }
\boldsymbol{h}^{m}(\mathcal{W}) = \sum_{i \in I} [\boldsymbol{h}(N) - \boldsymbol{h}(N \backslash i)] \quad \mathcal{W} \subseteq \mathcal{N}
\end{equation*}
Next $\boldsymbol{h}^{ti}$ was pushed onto the hyperplane such that $I(X_3,X_4) = 0$ and $I(X_1;X_2 | X_3,X_4) = 0$ through the linear mapping

\begin{eqnarray*}
\boldsymbol{h}_{AB} &&=  A_{34}B_{34,1} \boldsymbol{h}^{ti} \\
&&= \boldsymbol{h}^{ti} + (h^{ti}_3 + h^{ti}_4 - h^{ti}_{34})(\boldsymbol{r}^{3}_1 - \boldsymbol{r}^{\emptyset}_1) \\
&& + (h^{ti}_{123} + h^{ti}_{124} - h^{ti}_{34} - h^{ti}_{1234})(\boldsymbol{r}^{1}_2 - \boldsymbol{r}^{\emptyset}_3)
\end{eqnarray*}
After that, another linear mapping $C_{34}$ is used to further reduce the dimension of $G^{34}_{4}$ to 4.
{\small
\begin{eqnarray*}
&\boldsymbol{h}_{C} = C_{34}\boldsymbol{h}_{AB} = - Ingleton_{34}(\boldsymbol{h}^{ti}) \boldsymbol{f}_{34} \\
&+(h^{ti}_3 + h^{ti}_4 - h^{ti}_{34})\boldsymbol{r}^{\emptyset}_1 + (h^{ti}_{123} + h^{ti}_{124} - h^{ti}_{34} - h^{ti}_{1234}) \boldsymbol{r}^{\emptyset}_3 \\
&+\frac{1}{2}(h^{ti}_{13} + h^{ti}_{23} - h^{ti}_{3} - h^{ti}_{123} + h^{ti}_{14} + h^{ti}_{24} - h^{ti}_{4} - h^{ti}_{124}) (\boldsymbol{r}^{3}_1 + \boldsymbol{r}^{4}_1)\\
&+\frac{1}{2}(h^{ti}_{13} + h^{ti}_{14} - h^{ti}_{1} - h^{ti}_{134} + h^{ti}_{23} + h^{ti}_{24} - h^{ti}_{2} - h^{ti}_{234}) (\boldsymbol{r}^{1}_2 + \boldsymbol{r}^{2}_2)\\
&+\frac{1}{4}(h^{ti}_{12} + h^{ti}_{14} - h^{ti}_{1} - h^{ti}_{124} + h^{ti}_{12} + h^{ti}_{13} - h^{ti}_{1} - h^{ti}_{123} + \\
&h^{ti}_{12} + h^{ti}_{24} - h^{ti}_{2} - h^{ti}_{124} + h^{ti}_{12} + h^{ti}_{23} - h^{ti}_{2} - h^{ti}_{123} ) (\boldsymbol{r}^{13}_1 + \\
&\boldsymbol{r}^{14}_1+ \boldsymbol{r}^{23}_1 + \boldsymbol{r}^{24}_1)
\end{eqnarray*}
}

If we further normalize the last dimension to equal to one, by dividing through by it, the resulting polytope associated with the convex hull in the remaining three dimensions is three dimensional.  In order to coordinatize it, define $\boldsymbol{\alpha} = \frac{1}{4} \boldsymbol{f}_{34}$, $\boldsymbol{\beta} = \frac{1}{2}(\boldsymbol{r}^{3}_1 + \boldsymbol{r}^{4}_1)$, $\boldsymbol{\gamma} = \frac{1}{4} (\boldsymbol{r}^{1}_2 + \boldsymbol{r}^{2}_2)$ and $\boldsymbol{\delta} = \frac{1}{4} (\boldsymbol{r}^{13}_1 + \boldsymbol{r}^{14}_1 + \boldsymbol{r}^{23}_1 + \boldsymbol{r}^{24}_1)$, for any given transformed $\boldsymbol{g}$, we can write 
\begin{equation*}
\label{ }
\boldsymbol{g} = \bar{\alpha}_h \boldsymbol{\alpha} +  \bar{\beta}_h \boldsymbol{\beta} +  \bar{\gamma}_h \boldsymbol{\gamma} +  \bar{\delta}_h \boldsymbol{\delta}
\end{equation*} 
where $\bar{\alpha}_h + \bar{\beta}_h + \bar{\gamma}_h + \bar{\delta}_h = 1$. So in three dimensional space, we consider $\boldsymbol{\alpha}$ to be $(0, 0, 0)$, $\boldsymbol{\beta}$ to be $(\frac{1}{2}, \frac{\sqrt{3}}{2}, 0)$, $\boldsymbol{\gamma}$ to be $(1, 0, 0)$, $\boldsymbol{\delta}$ to be $(\frac{1}{2}, \frac{\sqrt{3}}{6}, \frac{\sqrt{6}}{3})$, so we can make the plot using $\bar{\beta}_h + \bar{\delta}_h$, $\bar{\gamma}_h + \bar{\delta}_h$  and $\bar{\alpha}_h$ as the three coordinate. 

\begin{figure}
\begin{center}
\includegraphics[width=.5\columnwidth]{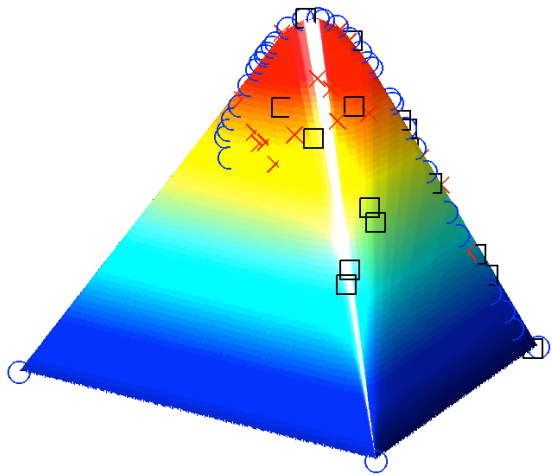}
\caption{The surface is a projection of a 3D face of the inner bound to $\bar{\Gamma}^*_4$ created with the numerical process described in the text with 4-atom distributions.  The blue circles are the from the entropic vectors from the extremal optimized four atom distributions, while the red $X$s and the black squares are the additional extremal optimized $\nAtoms$ distributions for $\nAtoms\in\{5,6\}$, respectively.}
\label{fg:inner4}
\end{center}
\end{figure}

See Figure \ref{fg:inner4} for a surface plot of the inner bound generated by $4$-atom supports, where we also plot the extremal points of $5$-atom(red X) and $6$-atom(black squares) inner bounds for comparison. Since we are transforming entropic vector from 15 dimension to 3 dimension, lots of extreme points of our 15 dimensional inner bound actually become redundant in this three dimensional space, so the number of points we can plot is significantly less than the number of extreme points we get from numerical optimization. As can be seen in Figure \ref{fg:inner4}, the extreme points of $4$-atom inner bound mostly lies in a curve, and there are some $5$-atom extreme points away from the convex hull generated with this curve, and some of $6$-atom extreme points can get even further away. 

\begin{figure}
\begin{center}
\includegraphics[width=.5\columnwidth]{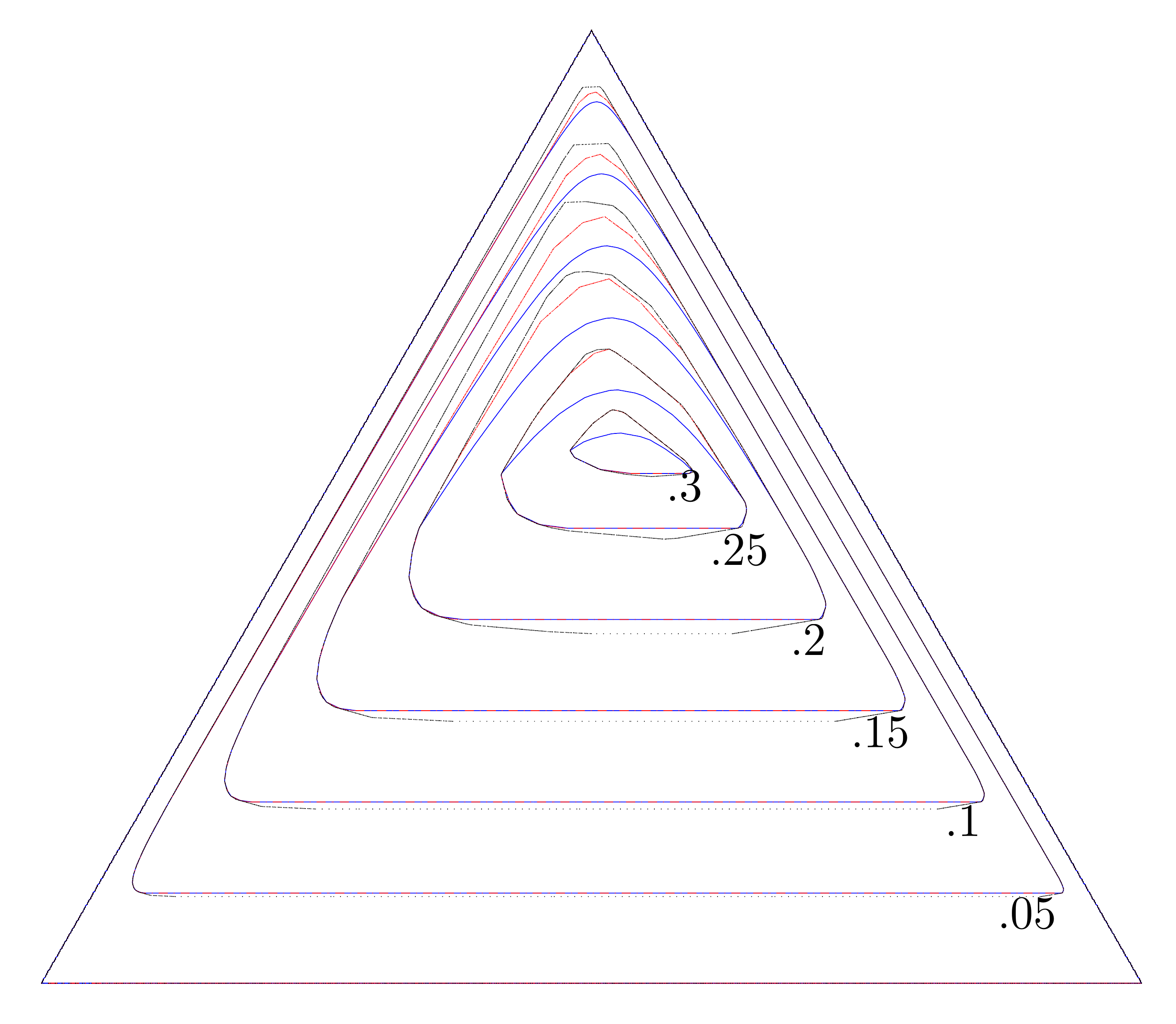}
\caption{Comparison of contour plots of inner bound created from $\leq \nAtoms$ atom distributions for $\nAtoms \in \{4,5,6\}$.  For each contour value, the inner most line is $\nAtoms=4$ while the outermost line is $\nAtoms=6$.  The numerical inner bounds generated from only four atom distributions are quite good in this space.}
\label{fg:innercontour}
\end{center}
\end{figure}

In order to better visualize the difference between the various inner bounds, we also compared the contour of inner bound generated by $\leq k$ atom supports for $k \in \{ 4, 5, 6 \}$, see Figure \ref{fg:innercontour} for this comparison plot where \emph{blue} line is $k = 4$, \emph{red} line is $k = 5$ and \emph{black} line is $k = 6$. As you can see, the contour is larger as more atoms are involved, meaning we can constantly get better inner bounds by increasing the number of atoms.  This is because as the number of atoms are increased, a greater and greater variety of distribution supports are found that can violate the Ingleton inequality. The increase of the inner bound from five to six atom is smaller than the increase from four to five atoms, which is consistent with the full dimensional volume calculation from Table \ref{tb:volume}.

\section{Perspectives on How to Use Information Geometry to Understand Entropy Geometry}\label{sec:infGeo}
The previous section utilized support enumeration followed by numerical optimization to map the unknown parts of the entropy.  While this did provide some numerical insight into which supports can have distributions that violate Ingleton and map to the unknown part of the entropy region, it did not give any analytical insight into how to parametrize probabilities so that the associated entropic vector would be in this unknown part, or be extremal.   Bearing this in mind, in this section we aim to study properties of probability distributions associated with entropic vectors that are extremal, in the sense that they give entropic vectors lying in faces of the Shannon outer bound, and also that violate Ingleton.  To do so, we will make use of information geometry, a discipline that in part studies ways to coordinatize probability distributions in a manner that relates properties such as independence to affine subspaces of the parameter space.


Information geometry, is a discipline in statistics which endows the manifold of probability distribution with a special ``dually flat'' differential geometric structure created by selecting a Riemannian metric based on the Fisher information and a family of affine connections called the $\alpha$-connections.

The natural divergence functions arising in information geometry, which include the Kullback Leibler divergence, can easily be related to the Shannon entropy by measuring the divergence with the uniform distribution.  Additionally, one could think of entropic vectors as characterizing simultaneous properties (entropies) of marginal distributions on subsets of variables, and this simultaneous marginal distribution relationship has a well studied structure in information geometry.

Building on these observations, here we wish to sketch out a couple of ideas in an attempt to show that characterizing the boundary of $\overline{\Gamma}^*_N$ could be thought of as a problem in information geometry, and that information geometry may in fact be a convenient framework to utilize to calculate these boundaries.  The perspective we will provide is by no means unique, however, the aim here is to provide a preliminary link between these disciplines and enough evidence about their relationship in the hope that we or other researchers in these areas may one day solve the apparently very difficult, yet fundamental and very important, problem of explicitly determining $\overline{\Gamma}^*_N$ by exploiting information geometric tools.

Along these lines, we will first review in a somewhat rough manner in \S \ref{ssec:infGeoRev} some concepts from information geometry related to entropic vectors.  Next, we will provide in \S \ref{ssec:infGeoSubMod} an information geometric characterization of those probability distributions associated with Shannon facets and some Shannon faces of the region of entropic vectors.  We then present an information geometric characterization of submodularity as a corollary.  Finally, we will close the paper in \S \ref{ssec:IngVio} with characterization of the 4 atom distributions (\ref{eq:4atoms}) which violate Ingleton via Information Geometry, and show a natural information geometric parameterization for them.


\subsection{Review of Some Relevant Ideas from Information Geometry}\label{ssec:infGeoRev}
Information geometry endows a manifold of probability distributions $p(x;\boldsymbol{\xi})$, parameterized by a vector of real numbers $\boldsymbol{\xi}=[\xi_i]$, with a Riemannian metric, or inner product between tangent vectors, given by the Fisher information:
\begin{equation}
g_{i,j}(\xi) \triangleq \mathbb{E}_\xi[ \frac{\partial \log p(x;\boldsymbol{\xi})}{\partial \xi_i} \frac{\partial \log p(x;\boldsymbol{\xi})}{\partial \xi_j} ]
\end{equation}
This allows us to calculate an inner product between two tangent vectors $ c = \sum_i c_i \partial_{\xi_i}  $ and $d= \sum_i d_i \partial_{\xi_i}  $ at a particular point $\xi$ in the manifold of probability distributions,  as
\begin{equation}
<c,d>_{\xi} = \sum_{i,j} c_i d_j g_{i,j}(\xi)
\end{equation}
Selecting this Riemannian metric (indeed, we could have selected others), and some additional structure, allows the differential geometric structure to be related to familiar properties of exponential families and mixture families of probability distributions.  The additional structure, called an affine connection, is given to the manifold to allow us establish a correspondence, called parallel translation, between tangent vectors living the tangent space at two different points along a curve in the manifold by solving a system of ordinary differential equations involving given functions on the manifold called connection coefficients.  Just as with a Riemannian metric, we have an infinite number possible of choices for the affine connection (embodied by selecting connection coefficients with respect to a particular parametrization), but if we choose particular ones, called the $\alpha$-connections, certain differential geometric notions like flatness and autoparallel submanifolds can be related to familiar notions of probability distribution families/subfamilies like exponential families and mixture families.  While the resulting theory is very elegant, it is also somewhat complex, and hence we must omit the general details, referring the interested reader to \cite{Amari_Book_04}, and only introduce a small subset of the concepts that can be utilized via a series of examples.

In particular, let's focus our attention on the manifold 
$\mathcal{O}(\boldsymbol{\mathcal{X}})$ 
of probability distributions for a random vector $\boldsymbol{X}=(X_1,\ldots,X_N)$ taking values on the Cartesian product $\boldsymbol{\mathcal{X}}^{\times} = \mathcal{X}_1\times \mathcal{X}_2 \times \cdots \times \mathcal{X}_N$.  
We already defined a $\prod_{n=1}^N | \mathcal{X}_n| -1$ dimensional vector  in (\ref{eq:eta}) as $\boldsymbol{\eta}$-coordinates, which we also call $m$-coordinate in information geometry. Alternatively we could parameterize the probability mass function for such a joint distribution with a vector $\boldsymbol{\theta}$, whose $\prod_{n=1}^N | \mathcal{X}_n| -1$ elements take the form
\begin{equation}
\label{thetacoordinatefirst}
\boldsymbol{\theta}= \left[ \log\left( \frac{p_{\mathbf{X}}(i_{1},\ldots,i_{N})}{ p_{\mathbf{X}}(1,\ldots,1)  } \right) \left| \begin{array}{c} i_k \in\{1,2,\ldots,|\mathcal{X}_k|\}, \\ k\in\{1,\ldots,N\}, \\ \sum^{N}_{k=1} i_k \neq N. \end{array} \right. \right]
\end{equation}
where $\mathcal{X}_n$ is a finite set with values denoted by $i_{k}$. These coordinates provide an alternate unique way of specifying the joint probability mass function $p_{\mathbf{X}}$, called the $e$-coordinates or $\boldsymbol{\theta}$ coordinates.

A subfamily of these probability mass functions associated with those $\boldsymbol{\eta}$ coordinates that take the form
\begin{equation}
\boldsymbol{\eta} = \mathbf{A} \boldsymbol{p} + \mathbf{b}
\end{equation}
for some $\boldsymbol{p}$ for any particular fixed $\mathbf{A}$ and $\mathbf{b}$, that is, that lie in an affine submanifold of the $\boldsymbol{\eta}$ coordinates, are said to form a $m$-\emph{autoparallel} submanifold of probability mass functions.  This is a not a definition, but rather a consequence of a theorem involving a great deal of additional structure which must omit here \cite{Amari_Book_04}.

Similarly, a subfamily of these probability mass functions associated with those $\boldsymbol{\theta}$ coordinates that take the form
\begin{equation}
\label{thetacoordinate}
\boldsymbol{\theta} = \mathbf{A} \boldsymbol{\lambda} + \mathbf{b}
\end{equation}
for some $\boldsymbol{\lambda}$ for any particular fixed $\mathbf{A}$ and $\mathbf{b}$, that is, that lie in an affine submanifold of the $\boldsymbol{\theta}$ coordinates, are said to form a $e$-\emph{autoparallel} submanifold of probability mass functions.

An e-autoparallel submanifold (resp. m-autoparallel submanifold) that is one dimensional, in that its $\boldsymbol{\lambda}$  (resp. $\boldsymbol{p})$ parameter vector is in fact a scalar, is called a $e$-\emph{geodesic} (resp. $m$-\emph{geodesic}).

On this manifold of probability mass functions for random variables taking values in the set $\boldsymbol{\mathcal{X}}$, we can also define the Kullback Leibler divergence, or relative entropy, measured in bits, according to
\begin{equation}
D(p_{\boldsymbol{X}} || q_{\boldsymbol{X}}) = \sum_{\boldsymbol{x}\in\boldsymbol{\mathcal{X}}} p_{\boldsymbol{X}}(\boldsymbol{x}) \log_2\left( \frac{ p_{\boldsymbol{X}}(\boldsymbol{x})}{ q_{\boldsymbol{X}}(\boldsymbol{x})} \right)
\end{equation}
Note that in this context $D(p||q) \geq 0$ with equality if and only if $p=q$, and hence this function is a bit like a distance, however it does \emph{not} in general satisfy symmetry or triangle inequality.

\begin{figure}
\centering
\includegraphics[width=.4\textwidth]{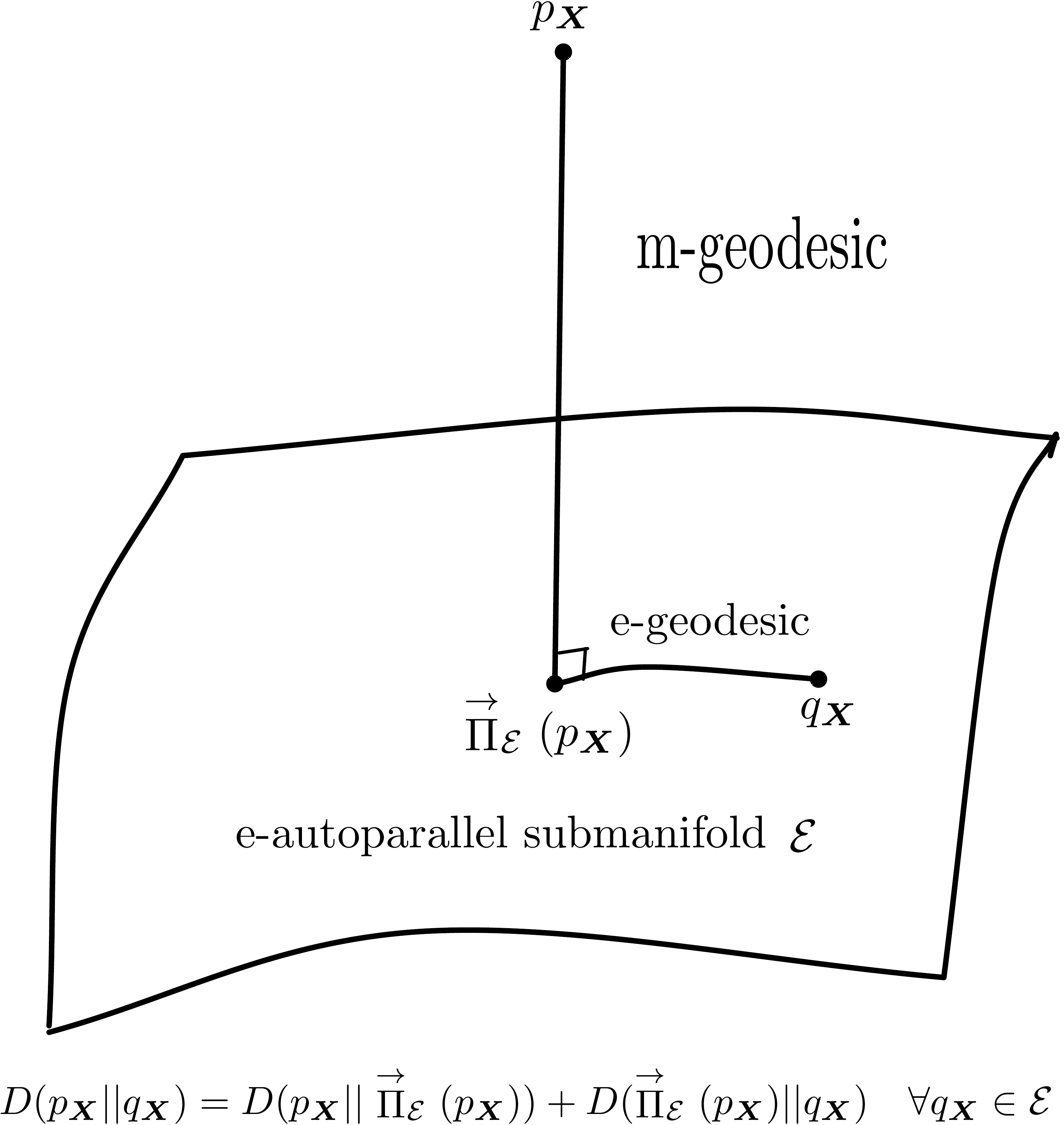}
\caption{Pythagorean style relation.}\label{fig:eproj}
\end{figure}

Let $\mathcal{E}$ be a particular e-autoparallel submanifold, and consider a probability distribution $p_{\boldsymbol{X}}$ not necessarily in this submanifold.  The problem of finding the point $\stackrel{\rightarrow}{\Pi}_{\mathcal{E}}(p_{\boldsymbol{X}})$ in $\mathcal{E}$ closest in Kullback Leibler divergence to $p_{\boldsymbol{X}}$ defined by
\begin{equation}
\stackrel{\rightarrow}{\Pi}_{\mathcal{E}}(p_{\boldsymbol{X}} )\triangleq \arg \min_{q_{\boldsymbol{X}} \in \mathcal{E}} D(p_{\boldsymbol{X}} || q_{\boldsymbol{X}})
\end{equation}
is well posed, and is characterized in the following two ways ( here $\stackrel{\rightarrow}{\Pi}_{\mathcal{E}}$ with a right arrow means we are minimizing over the second argument $q_{\boldsymbol{X}}$ ).  The tangent vector of the m-geodesic connecting $p_{\boldsymbol{X}}$ to $\stackrel{\rightarrow}{\Pi}_{\mathcal{E}}(p_{\boldsymbol{X}})$ is orthogonal, in the sense of achieving Riemannian metric value 0, at $\stackrel{\rightarrow}{\Pi}_{\mathcal{E}}(p_{\boldsymbol{X}})$ to the tangent vector of the e-geodesic connecting $\stackrel{\rightarrow}{\Pi}_{\mathcal{E}}(p_{\boldsymbol{X}})$ and any other point in $\mathcal{E}$.  Additionally, for any other point $q\in\mathcal{E}$, we have the Pythagorean like relation
\begin{equation}
D(p_{\boldsymbol{X}} || q_{\boldsymbol{X}}) = D(p_{\boldsymbol{X}} || \stackrel{\rightarrow}{\Pi}_{\mathcal{E}}(p_{\boldsymbol{X}}) ) + D( \stackrel{\rightarrow}{\Pi}_{\mathcal{E}}(p_{\boldsymbol{X}}) || q_{\boldsymbol{X}}).
\end{equation}
This relationship, which is an important one in information geometry \cite{Amari_Book_04}, is depicted in Fig. \ref{fig:eproj}.

\subsection{Information Geometric Structure of the Shannon Faces of the Region of Entropic Vectors}\label{ssec:infGeoSubMod}

\begin{figure}
\centering
\includegraphics[width=.4\textwidth]{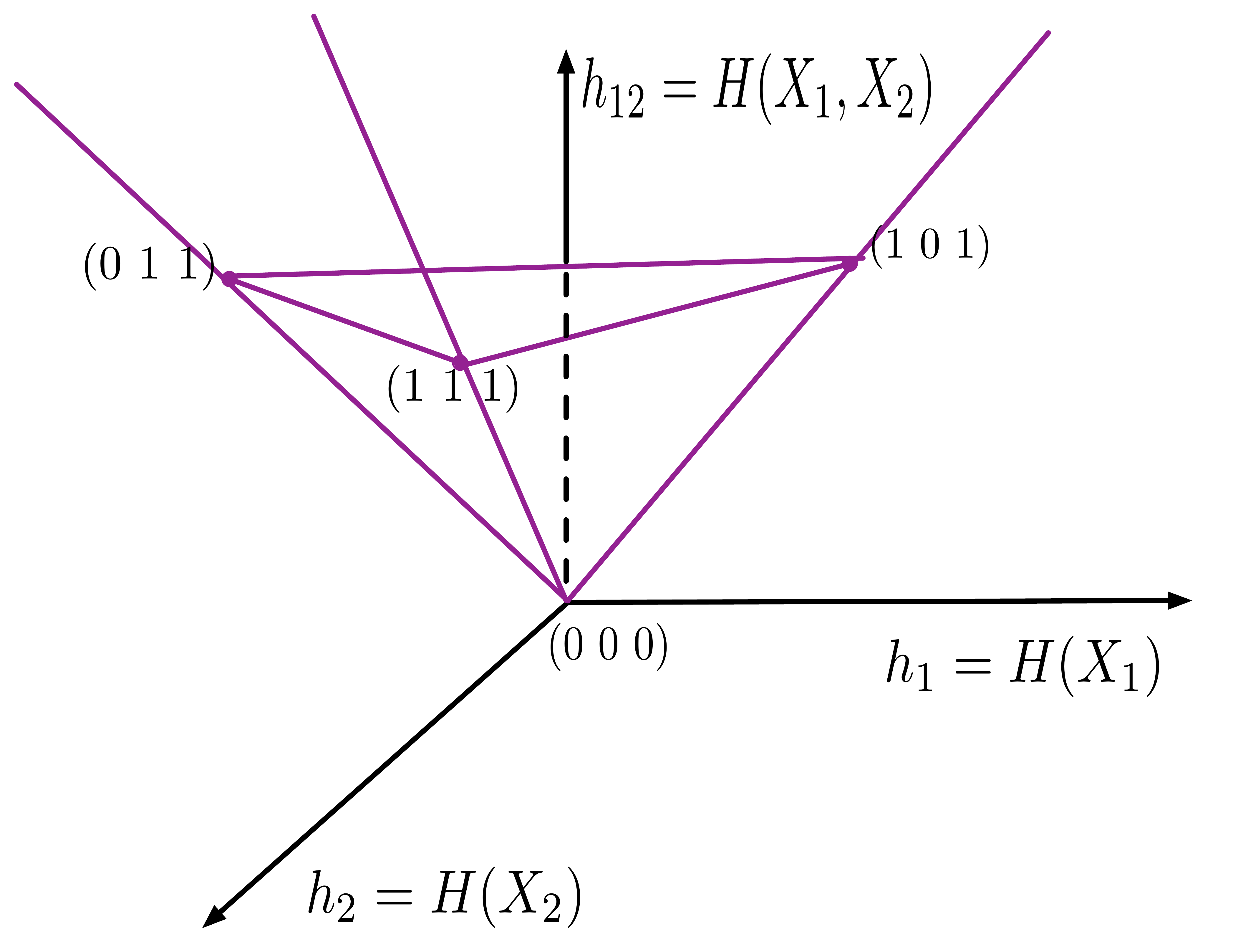}
\caption{Region of entropic vector $\Gamma^*_2$.}\label{fig:region1}
\end{figure}

As identified in \S \ref{ssec:boundsbest}, $\Gamma_2 = \Gamma^*_2$ and $\Gamma_3 = \overline{\Gamma}^*_3$, implying that $\Gamma^*_2$ and $\overline{\Gamma}^*_3$ are fully characterized by Shannon type information inequalities. For example, when N = 2,  $\Gamma^{*}_2$ is a 3-dimensional polyhedral cone characterized by $H(X_{1},X_{2}) - H(X_{1}) \geqslant 0$, $H(X_{1},X_{2}) - H(X_{2}) \geqslant 0$ and $I(X_{1};X_{2}) = H(X_{1})+H(X_{2})-H(X_{1},X_{2}) \geqslant 0$ as depicted in Fig. \ref{fig:region1}.

For $N=4$, even though $\Gamma_4 \neq \overline{\Gamma}^*_4$ and the region is not a polyhedral cone, there are still many exposed faces of $\overline{\Gamma}^*_N$ defined by attaining equality in a particular Shannon type information inequality of the form (\ref{eq:subMod}) or (\ref{eq:nonDec}).  Such exposed faces of $\overline{\Gamma}^*_N$ could be referred to as the ``Shannon facets'' of entropy, and in this section we will first aim to characterize the distributions associated with these Shannon facets via information geometry. \\

Let $\mathcal{E}_{\mathcal{A}}^{\perp}$ be the submanifold of probability distributions for which $\boldsymbol{X}_{\mathcal{A}}$ and $\boldsymbol{X}_{\mathcal{A}^c} = \boldsymbol{X}_{\mathcal{N}\setminus \mathcal{A}}$ are independent
\begin{equation}
\mathcal{E}_{\mathcal{A}}^{\perp} = \left\{p_{\boldsymbol{X}}(\cdot) \left|\ p_{\boldsymbol{X}}(\boldsymbol{x})= p_{\boldsymbol{X}_{\mathcal{A}}}(\boldsymbol{x}_{\mathcal{A}}) p_{\boldsymbol{X}_{\mathcal{A}^c}}(\boldsymbol{x}_{\mathcal{A}^c}) \quad \forall \boldsymbol{x} \right. \right\}
\end{equation}
then we define $\mathcal{E}_{\mathcal{A}\cup \mathcal{B}}^{\perp} $,  $\mathcal{E}_{\mathcal{A},\mathcal{B}}^{\leftrightarrow,\perp}$ and $\mathcal{M}_{\mathcal{A}}$ as follows
\begin{eqnarray*}
&&\mathcal{E}_{\mathcal{A}\cup \mathcal{B}}^{\perp} =  \left\{  p_{\boldsymbol{X}}(\cdot) \left|\ p_{\boldsymbol{X}}= p_{\boldsymbol{X}_{(\mathcal{A}\cup \mathcal{B})}} p_{\boldsymbol{X}_{(\mathcal{A}\cup \mathcal{B})^c}} \right.   \right\} \\
&&\mathcal{E}_{\mathcal{A},\mathcal{B}}^{\leftrightarrow,\perp} = \left\{  p_{\boldsymbol{X}}(\cdot) \left|\ p_{\boldsymbol{X}}= p_{\boldsymbol{X}_{\mathcal{A}}  } p_{\boldsymbol{X}_{\mathcal{B}\setminus\mathcal{A}} | \boldsymbol{X}_{\mathcal{A}\cap\mathcal{B}}  }  p_{\boldsymbol{X}_{(\mathcal{A}\cup \mathcal{B})^c}} \right. \right\} \\
&&\mathcal{M}_{\mathcal{A}}^w =\left\{  p_{\boldsymbol{X}}(\cdot) \left|\ p_{\boldsymbol{X}} = p_{\boldsymbol{X}_{\mathcal{A}}  } \cdot \delta_{\mathcal{A}^c}^w \right. \right\} \\
\end{eqnarray*}
where $\delta_{\mathcal{A}^c}^w =
\left \{
  \begin{array}{l l}
    1 & \quad \text{if}\ \boldsymbol{X}_{\mathcal{A}^c} = w(\boldsymbol{X}_{\mathcal{A}}) \\
    0 & \quad \text{otherwise}\\
  \end{array}
  \right. $ for some fixed function $w:\boldsymbol{\mathcal{X}}_{\mathcal{A}} \rightarrow \boldsymbol{\mathcal{X}}_{\mathcal{A}^c}$. 
  Note $\mathcal{E}_{\mathcal{A},\mathcal{B}}^{\leftrightarrow,\perp}$ is a submanifold of $\mathcal{E}_{\mathcal{A}\cup \mathcal{B}}^{\perp} $ such that the random variables $\boldsymbol{X}_{\mathcal{A} \cup \mathcal{B}}$, in addition to being independent from $\boldsymbol{X}_{(\mathcal{A}\cup \mathcal{B})^c}$ form the Markov chain $ \boldsymbol{X}_{\mathcal{A}\setminus \mathcal{B}} \leftrightarrow \boldsymbol{X}_{\mathcal{A} \cap \mathcal{B}} \leftrightarrow \boldsymbol{X}_{\mathcal{B} \setminus \mathcal{A}} $.
These sets of distributions will be useful because $I (\boldsymbol{X}_{\mathcal{A} \cup \mathcal{B}};\boldsymbol{X}_{(\mathcal{A} \cup \mathcal{B})^c}) =  h_{\mathcal{A} \cup \mathcal{B}} + h_{(\mathcal{A} \cup \mathcal{B})^c}-h_{\mathcal{N}} = 0$ for every distribution in $\mathcal{E}_{\mathcal{A}\cup \mathcal{B}}^{\perp}$ and $\mathcal{E}_{\mathcal{A},\mathcal{B}}^{\leftrightarrow,\perp}$,
$I(\boldsymbol{X}_{\mathcal{A}\setminus\mathcal{B}} ; \boldsymbol{X}_{\mathcal{B}\setminus\mathcal{A}}  | \boldsymbol{X}_{\mathcal{A} \cap \mathcal{B}}) = h_{\mathcal{A}} + h_{\mathcal{B}} - h_{(\mathcal{A} \cap \mathcal{B})} - h_{(\mathcal{A} \cup \mathcal{B})} = 0$ for every distribution in $\mathcal{E}_{\mathcal{A},\mathcal{B}}^{\leftrightarrow,\perp}$, and
$h_{\mathcal{N}} - h_{\mathcal{A}} = 0$ for every distribution in $\mathcal{M}_{\mathcal{A}}^w$ for any $w$. \\
\indent We will show that the sets of distributions $\mathcal{E}_{\mathcal{A}\cup \mathcal{B}}^{\perp} $,  $\mathcal{E}_{\mathcal{A},\mathcal{B}}^{\leftrightarrow,\perp}$, $\mathcal{M}_{\mathcal{A}}^w$ are of interest because they correspond to every Shannon facet.  This is because every facet of $\Gamma_N$, which must correspond to equality in an inequality of the form (\ref{eq:subMod}) or (\ref{eq:nonDec}), can be regarded as some conditional entropy, mutual information, or conditional mutual information being identically zero.
For example, as depicted in Fig. \ref{fig:region1}, $\Gamma^{*}_2$ has three facets corresponding to set of distributions $\mathcal{E}_{1\cup 2}^{\perp} $, $\{\mathcal{M}_{1}^w\}$ and $\{\mathcal{M}_{2}^w\}$ respectively.   Surprisingly, while the entropy vector is a nonlinear function of the joint distribution, and these families of distributions correspond to the intersection of affine sets with the region of entropic vectors, they are themselves, when regarded with the correct information geometric parameterization, \emph{associated with affine sets of distributions}.\\

In order to prove the information geometric properties of Shannon facets, we first give the equivalent condition of $I(\boldsymbol{X}_{\mathcal{A}\setminus\mathcal{B}} ; \boldsymbol{X}_{\mathcal{B}\setminus\mathcal{A}}  | \boldsymbol{X}_{\mathcal{A} \cap \mathcal{B}}) = 0$ in distribution space, especially in $\boldsymbol{\theta}$ coordinate as mentioned in (\ref{thetacoordinate}). 

\begin{lemma} \label{CIThetaSpace}
Let $\boldsymbol{X} = \{ X_1, X_2, \cdots, X_N\}$, $\mathcal{A},\mathcal{B} \subset \mathcal{N} = \{1,2,\cdots N\}$ where $|\mathcal{X}_{\mathcal{A}\setminus\mathcal{B}}| = m$, $|\mathcal{X}_{\mathcal{B}\setminus\mathcal{A}}| = n$ and $|\mathcal{X}_{\mathcal{A} \cap \mathcal{B}}| = q$.
For each assignment $C_0$ of $\boldsymbol{X}_{\mathcal{A} \cap \mathcal{B}}$, define the following equations for $i  = 1,2, \cdots, m-1$ and $j = 1,2, \cdots, n-1$ for the probabilities $p_{\mathbf{X}_{\mathcal{A}\setminus \mathcal{B}},\mathbf{X}_{\mathcal{B}\setminus \mathcal{A}},\mathcal{X}_{\mathcal{A}\cap\mathcal{B}}}$:
\begin{equation}
\label{probmulti}
\frac{p_{0jC_0} p_{i0C_0}}{p_{00C_0}p_{ijC_0}} = 1
\end{equation}
which is equivalent with hyperplane (\ref{thetamulti}) in $\boldsymbol{\theta}$ coordinates for $p_{\mathbf{X}_{\mathcal{A}\cup\mathcal{B}}}$,
\begin{equation}
\label{thetamulti}
\theta_{0jC_0} + \theta_{i0C_0} - \mathds{1}[ C_0 \neq \textbf{0}] \cdot\theta_{00C_0} -\theta_{ijC_0} = 0
\end{equation}
then $I(\boldsymbol{X}_{\mathcal{A}\setminus\mathcal{B}} ; \boldsymbol{X}_{\mathcal{B}\setminus\mathcal{A}}  | \boldsymbol{X}_{\mathcal{A} \cap \mathcal{B}}) = 0$ and the set of equations (\ref{probmulti}), (\ref{thetamulti}) are equivalent.
\end{lemma}
\textbf{Proof}: 
From Equation (\ref{probmulti}), it can be verified that
\begin{equation}
\label{probRedund}
\frac{p_{wjC_0} p_{ieC_0}}{p_{weC_0}p_{ijC_0}} = 1
\end{equation}
for any distinct $i$, $j$, $w$ and $e$ such that $w, i \in \{0,1, \cdots, m-1\}$ and $j,e \in \{0,1,\cdots, n-1\}$. More specifically, the equations in (\ref{probRedund}) but not in (\ref{probmulti}) all can be derived by combining some of the $(m-1)(n-1)$ equations in (\ref{probmulti}).

$\Rightarrow$ Since $I(\boldsymbol{X}_{\mathcal{A}\setminus\mathcal{B}} ; \boldsymbol{X}_{\mathcal{B}\setminus\mathcal{A}}  | \boldsymbol{X}_{\mathcal{A} \cap \mathcal{B}}) = 0$, for each assignment $C_0$ of $\boldsymbol{X}_{\mathcal{A} \cap \mathcal{B}}$, it can be verified \cite{Yeung_ITNETCODBOOK_08} that 
\begin{eqnarray*}
p_{     \boldsymbol{X_{\mathcal{A} \setminus \mathcal{B}}}          \boldsymbol{X_{\mathcal{B}\setminus \mathcal{A}}}       |        \boldsymbol{X_{\mathcal{A} \cap \mathcal{B}}}    }      (x_{\mathcal{A}\setminus\mathcal{B}} = i, x_{\mathcal{B}\setminus\mathcal{A}} = j, x_{\mathcal{A} \cap \mathcal{B}} = C_0) \\
= p_{\boldsymbol{X_{\mathcal{A} \setminus \mathcal{B}}} | \boldsymbol{X_{\mathcal{A} \cap \mathcal{B}}}}(x_{\mathcal{A}\setminus\mathcal{B}} = i | x_{\mathcal{A} \cap \mathcal{B}} = C_0) \\
p_{\boldsymbol{X_{\mathcal{B}\setminus \mathcal{A}}}| \boldsymbol{X_{\mathcal{A} \cap \mathcal{B}}}}(x_{\mathcal{B}\setminus\mathcal{A}} = j | x_{\mathcal{A} \cap \mathcal{B}} = C_0)
\end{eqnarray*}
which make sure the nominator and denominator cancelled with each other in (\ref{probmulti}) and (\ref{probRedund}). 

$\Leftarrow$ Now suppose (\ref{probmulti}) , (\ref{thetamulti}) and (\ref{probRedund}) hold for $w, i \in \{0,1, \cdots, m-1\}$ and $j,e \in \{0,1,\cdots, n-1\}$, it is suffices to show for each assignment $C_0$ of $\boldsymbol{X}_{\mathcal{A} \cap \mathcal{B}}$,
{\small
\begin{equation}
\label{mutual_zero}
H(\boldsymbol{X_{\mathcal{A} \setminus \mathcal{B}}} |  \boldsymbol{X_{\mathcal{A} \cap \mathcal{B}}}) + H(\boldsymbol{X_{\mathcal{B}\setminus \mathcal{A}}} |  \boldsymbol{X_{\mathcal{A} \cap \mathcal{B}}} ) = H(\boldsymbol{X_{\mathcal{A} \setminus \mathcal{B}}}, \boldsymbol{X_{\mathcal{B}\setminus \mathcal{A}}} |  \boldsymbol{X_{\mathcal{A} \cap \mathcal{B}}} )
\end{equation}
}

We can calculate:
\begin{equation}
\label{h1}
H(\boldsymbol{X_{\mathcal{A} \setminus \mathcal{B}}} | \boldsymbol{X_{\mathcal{A} \cap \mathcal{B}}} = C_0)  = \sum_{i = 0}^{m-1} [(\sum_{j = 0}^{n-1}p_{i j C_0}) \log(\sum_{j = 0}^{n-1}p_{i j C_0}) ]
\end{equation}

\begin{equation}
\label{h2}
H(\boldsymbol{X_{\mathcal{B}\setminus \mathcal{A}}} | \boldsymbol{X_{\mathcal{A} \cap \mathcal{B}}} = C_0)  = \sum_{j = 0}^{n-1} [(\sum_{i = 0}^{m-1}p_{i j C_0}) \log(\sum_{i = 0}^{m-1}p_{i j C_0}) ]
\end{equation}

\begin{equation}
\label{h12}
H(\boldsymbol{X_{\mathcal{A} \setminus \mathcal{B}}}, \boldsymbol{X_{\mathcal{B}\setminus \mathcal{A}}} | \boldsymbol{X_{\mathcal{A} \cap \mathcal{B}}} = C_0)  = \sum_{i = 0}^{m-1} \sum_{j = 0}^{n-1}p_{i j C_0} \log p_{i j C_0} 
\end{equation}

From (\ref{h1}) and (\ref{h2}) we have
\begin{eqnarray}
\label{h1plus2}
H(\boldsymbol{X_{\mathcal{A} \setminus \mathcal{B}}} | \boldsymbol{X_{\mathcal{A} \cap \mathcal{B}}} = C_0) + H(\boldsymbol{X_{\mathcal{B}\setminus \mathcal{A}}} | \boldsymbol{X_{\mathcal{A} \cap \mathcal{B}}} = C_0)  = \\ \nonumber
\sum_{i = 0}^{m-1} \sum_{j = 0}^{n-1}p_{i j C_0} \log [(\sum_{e = 0}^{n-1} p_{i e C_0})(\sum_{w = 0}^{m-1} p_{w j C_0})]
\end{eqnarray}
where $(\sum_{e = 0}^{n-1} p_{i e C_0})(\sum_{w = 0}^{m-1} p_{w j C_0}) = p_{i j C_0}$ when (\ref{probRedund}) are satisfied, in which case (\ref{mutual_zero}) hold.
\hspace*{\fill} $\Box$\\

Now let's take $I(X_1;X_2|X_3) = 0$ as a example:
\begin{example} \label{CIThetaSpace_example} Let $\boldsymbol{X} = \{ X_1, X_2, X_3\}$, where $|\mathcal{X}_1| = m$, $|\mathcal{X}_2| = n$ and $|\mathcal{X}_3| = q$. For $mnq$ different values of $\boldsymbol{X}$, if we want to parameterize this manifold $\mathcal{O}_3$ in $\boldsymbol{\theta}$ coordinate, we will need $(mnq-1)$ different parameters: $\theta_1$, $\theta_2$, $\cdots$, $\theta_{mnq-1}$, which can be defined as in (\ref{thetacoordinatefirst}).  We define submanifold $\mathcal{E}_3$ to satisfy the constraint $I(X_1;X_2|X_3) = 0$. It is easy to verify that the dimension of $\mathcal{O}_3$ is $(mnq-1)$, the dimension of $\mathcal{E}_3$ is $((m+n-1)q-1)$. 
We know any hyperplane $\sum_i^{mnq-1} \lambda_i \theta_i = 0$ is a $(mnq-2)$ dimensional e-autoparallel submanifold in $\mathcal{O}_3$. Then subtracting $((m+n-1)q-1) + 1$ from $mnq-2$, we know $\mathcal{E}_3$ is the intersection of $(m-1)(n-1)q$ different hyperplanes, which are $\sum_{i=1}^{mnq-1}\lambda_i^{k}\theta_i = 0$ for $k =1, 2,  \cdots, (m-1)(n-1)q$. For $i  = 1,2, \cdots, m-1$, $j = 1,2, \cdots, n-1$ and $r = 1,2, \cdots, q$, each hyperplane $\sum_{i=1}^{mnq-1}\lambda_i^{k}\theta_i = 0$ corresponding to a constraint on the probability:
\begin{equation}
\label{probsingle}
\frac{p_{0jr} p_{i0r}}{p_{00r}p_{ijr}} = 1
\end{equation}
which can be written as
\begin{equation}
\label{thetasingle}
\theta_{0jr} + \theta_{i0r} - \mathds{1}[ r \neq 0] \cdot\theta_{00r} -\theta_{ijr} = 0
\end{equation}
where $\mathds{1}[ r \neq 0] = 0$ when $r = 0$. 
\end{example}

Now we can use Lemma \ref{CIThetaSpace} to prove Theorem \ref{ig_lemma1}, then use the relationship between $m$-autoparallel submanifold and affine subspace to prove Theorem \ref{ig_lemma2}:
\begin{theorem} \label{ig_lemma1}
$\mathcal{E}_{\mathcal{A},\mathcal{B}}^{\leftrightarrow,\perp} \subseteq \mathcal{E}_{\mathcal{A}\cup \mathcal{B}}^{\perp}  \subseteq \mathcal{O}(\boldsymbol{\mathcal{X}})$,
$\mathcal{E}_{\mathcal{A},\mathcal{B}}^{\leftrightarrow,\perp}$ is an e-autoparallel submanifold of $\mathcal{E}_{\mathcal{A}\cup \mathcal{B}}^{\perp}$ and $\mathcal{E}_{\mathcal{A}\cup \mathcal{B}}^{\perp}$ is an e-autoparallel submanifold of $\mathcal{O}(\boldsymbol{\mathcal{X}})$.
\end{theorem}
\noindent \textbf{Proof}:
Follows directly from Bayes' rule,  we can rewrite $\mathcal{E}_{\mathcal{A}\cup \mathcal{B}}^{\perp} $ and $\mathcal{O}(\boldsymbol{\mathcal{X}})$ as
\begin{eqnarray*}
&&\mathcal{E}_{\mathcal{A}\cup \mathcal{B}}^{\perp} =\left\{  p_{\boldsymbol{X}}(\cdot) \left|\ p_{\boldsymbol{X}}= p_{\boldsymbol{X}_{\mathcal{A}}  } p_{\boldsymbol{X}_{\mathcal{B}\setminus\mathcal{A}} | \boldsymbol{X}_{\mathcal{A}}  }  p_{\boldsymbol{X}_{(\mathcal{A}\cup \mathcal{B})^c}} \right. \right\} \\
&&\mathcal{O}(\boldsymbol{\mathcal{X}}) =  \left\{  p_{\boldsymbol{X}}(\cdot) \left|\ p_{\boldsymbol{X}}= p_{\boldsymbol{X}_{(\mathcal{A}\cup \mathcal{B})}} p_{\boldsymbol{X}_{(\mathcal{A}\cup \mathcal{B})^c | (\mathcal{A}\cup \mathcal{B})}} \right.   \right\}
\end{eqnarray*}
then we can easily verify $\mathcal{E}_{\mathcal{A},\mathcal{B}}^{\leftrightarrow,\perp} \subseteq \mathcal{E}_{\mathcal{A}\cup \mathcal{B}}^{\perp}  \subseteq \mathcal{O}(\boldsymbol{\mathcal{X}})$.
Since from Lemma \ref{CIThetaSpace}, the equivalent condition of $I(\boldsymbol{X}_{\mathcal{A}\setminus\mathcal{B}} ; \boldsymbol{X}_{\mathcal{B}\setminus\mathcal{A}}  | \boldsymbol{X}_{\mathcal{A} \cap \mathcal{B}})  = 0$ in $e$-coordinate for $p_{\mathbf{X}_{\mathcal{A}\cup\mathcal{B}}}$ is the intersection of a sequences of hyperplanes (\ref{thetamulti}) and the case of $\mathcal{A} \cap \mathcal{B} = \emptyset$ can be considered as a special case of Lemma \ref{CIThetaSpace}.
Then from the definition of autoparallel submanifold, $\mathcal{E}_{\mathcal{A},\mathcal{B}}^{\leftrightarrow,\perp}$ is e-autoparallel in $\mathcal{E}_{\mathcal{A}\cup \mathcal{B}}^{\perp}$, and $\mathcal{E}_{\mathcal{A}\cup \mathcal{B}}^{\perp}$ is e-autoparallel in $\mathcal{O}(\boldsymbol{\mathcal{X}})$.
\hspace*{\fill} $\Box$ \\
\begin{theorem} \label{ig_lemma2}
Let $\mathcal{A}\subseteq \mathcal{B} \subseteq \mathcal{N} = \{1,2,\cdots N\}$,  
then $\mathcal{M}_{\mathcal{A}}^{w} \subseteq \mathcal{M}_{\mathcal{B}}^{w'} \subseteq \mathcal{O}(\boldsymbol{\mathcal{X}})$, $\mathcal{M}_{\mathcal{A}}^{w}$ is a m-autoparallel submanifold of $\mathcal{M}_{\mathcal{B}}^{w'}$ and $\mathcal{M}_{\mathcal{B}}^{w'}$ is a m-autoparallel submanifold of $\mathcal{O}(\boldsymbol{\mathcal{X}})$.
\end{theorem}
\noindent \textbf{Proof}:
It follows directly from the definition of m-autoparallel submanifold since for the m-affine coordinate of $\mathcal{O}(\boldsymbol{\mathcal{X}})$ and $\mathcal{M}_{\mathcal{B}}^{w'}$, we can easily find the corresponding matrix $A$ and vector $B$ such that $\mathcal{M}_{\mathcal{B}}^{w'}$ is  affine subspace of $\mathcal{O}(\boldsymbol{\mathcal{X}})$; similarly for the m-affine coordinate of $\mathcal{M}_{\mathcal{A}}^w$ and $\mathcal{M}_{\mathcal{B}}^{w'}$, we can easily find the corresponding matrix $\mathbf{A}$ and vector $\mathbf{b}$ such that $\mathcal{M}_{\mathcal{A}}^w$ is  affine subspace of $\mathcal{M}_{\mathcal{B}}^w$.
\hspace*{\fill} $\Box$
\\

\begin{figure}
\centering
\includegraphics[width=.47\textwidth]{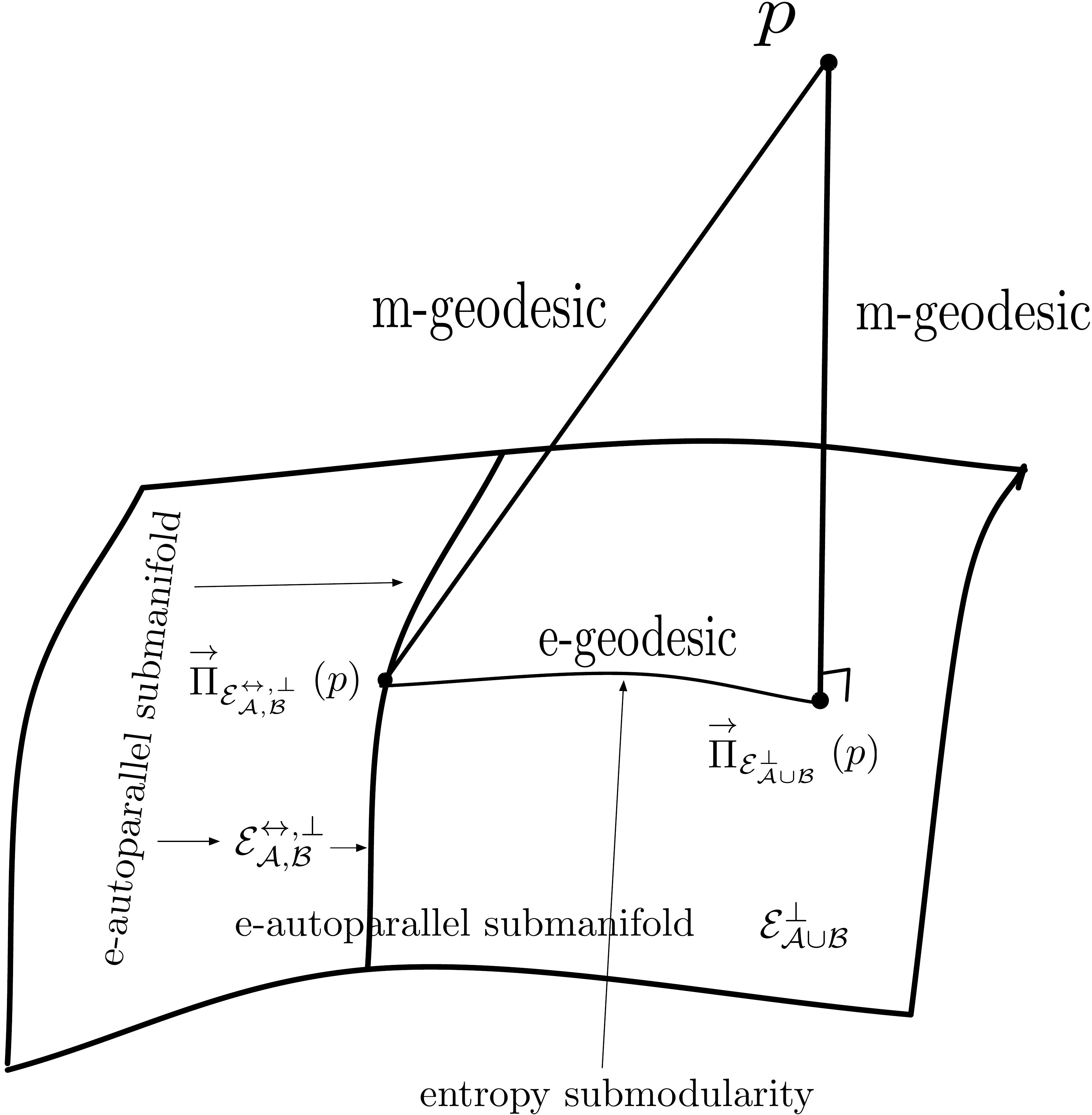}
\caption{Submodularity of the entropy function is equivalent to the non-negativity of a divergence $D(\stackrel{\rightarrow}{\pi}_{\mathcal{E}_{\mathcal{A}\cup\mathcal{B}}^{\perp}}(p) || \stackrel{\rightarrow}{\pi}_{\mathcal{E}_{\mathcal{A},\mathcal{B}}^{\leftrightarrow,\perp}}(p) )$  between two information projections, one projection ($\stackrel{\rightarrow}{\pi}_{\mathcal{E}_{\mathcal{A},\mathcal{B}}^{\leftrightarrow,\perp}}(p)$) is to a set that is a submanifold the other projection's ($\stackrel{\rightarrow}{\pi}_{\mathcal{E}_{\mathcal{A}\cup\mathcal{B}}^{\perp}}(p)$) set.  A Pythagorean style relation shows that for such an arrangement $D(p|| \stackrel{\rightarrow}{\pi}_{\mathcal{E}_{\mathcal{A},\mathcal{B}}^{\leftrightarrow,\perp}}(p) ) = D(p || \stackrel{\rightarrow}{\pi}_{\mathcal{E}_{\mathcal{A}\cup\mathcal{B}}^{\perp}}(p) ) + D(\stackrel{\rightarrow}{\pi}_{\mathcal{E}_{\mathcal{A}\cup\mathcal{B}}^{\perp}}(p) || \stackrel{\rightarrow}{\pi}_{\mathcal{E}_{\mathcal{A},\mathcal{B}}^{\leftrightarrow,\perp}}(p) )$.}\label{fig:pithy}
\end{figure}

Theorem \ref{ig_lemma1} and Theorem \ref{ig_lemma2}, have shown that Shannon facets are associated with affine subsets of the family of probability distribution, when it is regarded in an appropriate parameterization.  In fact, as we shall presently show, all Shannon Type information inequalities correspond to the positivity of divergences of projections to these submanifolds. Note that $\mathcal{E}^{\leftrightarrow,\perp}_{\mathcal{A},\mathcal{B}}$ is set up so that the difference between the right and left sides in the entropy submodularity inequality (\ref{eq:subMod}) is zero for every probability distribution in $\mathcal{E}^{\leftrightarrow,\perp}_{\mathcal{A},\mathcal{B}}$.  The nested nature of these e-autoparallel submanifolds, and the associated Pythagorean relation, is one way to view the submodularity of entropy, as we shall now explain with Figure \ref{fig:pithy} and Corollary \ref{thm:proj}.
\begin{corollary} \label{thm:proj}
The submodularity (\ref{eq:subMod}) of the entropy function can be viewed as a consequence of Pythagorean style information projection relationships depicted in Figure \ref{fig:pithy}.  In particular, submodularity is equivalent to the inequality
\begin{equation}
D\left(\stackrel{\rightarrow}{\Pi}_{\mathcal{E}_{\mathcal{A}\cup\mathcal{B}}^{\perp}}(p_{\boldsymbol{X}}) || \stackrel{\rightarrow}{\Pi}_{\mathcal{E}_{\mathcal{A},\mathcal{B}}^{\leftrightarrow,\perp}}(p_{\boldsymbol{X}}) \right) \geq 0
\end{equation}
since
\begin{eqnarray*}
D\left(\stackrel{\rightarrow}{\Pi}_{\mathcal{E}_{\mathcal{A}\cup\mathcal{B}}^{\perp}}(p_{\boldsymbol{X}}) || \stackrel{\rightarrow}{\Pi}_{\mathcal{E}_{\mathcal{A},\mathcal{B}}^{\leftrightarrow,\perp}}(p_{\boldsymbol{X}}) \right) = H(\boldsymbol{X}_{\mathcal{A}}) + H(\boldsymbol{X}_{\mathcal{B}}) \\-H(\boldsymbol{X}_{\mathcal{A}\cap\mathcal{B}}) - H(\boldsymbol{X}_{\mathcal{A}\cup\mathcal{B}})
\end{eqnarray*}
\end{corollary}
\noindent \textbf{Proof}: The projections are
\begin{equation}\label{eq:projRes1}
\stackrel{\rightarrow}{\Pi}_{\mathcal{E}_{\mathcal{A}\cup\mathcal{B}}^{\perp}}(p_{\boldsymbol{X}}) = p_{\boldsymbol{X}_{\mathcal{A}\cup\mathcal{B}}} p_{\boldsymbol{X}_{(\mathcal{A}\cup\mathcal{B})^c}}
\end{equation}
and
\begin{equation}\label{eq:projRes2}
 \stackrel{\rightarrow}{\Pi}_{\mathcal{E}_{\mathcal{A},\mathcal{B}}^{\leftrightarrow,\perp}}(p_{\boldsymbol{X}}) = p_{\boldsymbol{X}_{\mathcal{A}\setminus \mathcal{B}} | \boldsymbol{X}_{\mathcal{A}\cap \mathcal{B}}} p_{\boldsymbol{X}_{\mathcal{B}}} p_{\boldsymbol{X}_{(\mathcal{A}\cup\mathcal{B})^c}}
\end{equation}
since for $\stackrel{\rightarrow}{\Pi}_{\mathcal{E}_{\mathcal{A}\cup\mathcal{B}}^{\perp}}(p_{\boldsymbol{X}})$ given by (\ref{eq:projRes1})  for every $q_{\boldsymbol{X}}\in\mathcal{E}_{\mathcal{A}\cup\mathcal{B}}^{\perp}$ we can reorganize the divergence as
\begin{eqnarray*}
D(p_{\boldsymbol{X}} || q_{\boldsymbol{X}}) = D(p_{\boldsymbol{X}} || \stackrel{\rightarrow}{\Pi}_{\mathcal{E}_{\mathcal{A}\cup\mathcal{B}}^{\perp}}(p_{\boldsymbol{X}})) \\+ D(\stackrel{\rightarrow}{\Pi}_{\mathcal{E}_{\mathcal{A}\cup\mathcal{B}}^{\perp}}(p_{\boldsymbol{X}}) || q_{\boldsymbol{X}_{(\mathcal{A}\cup\mathcal{B})} }q_{\boldsymbol{X}_{(\mathcal{A}\cup\mathcal{B})^c} })
\end{eqnarray*}
and for $\stackrel{\rightarrow}{\Pi}_{\mathcal{E}_{\mathcal{A},\mathcal{B}}^{\leftrightarrow,\perp}}(p_{\boldsymbol{X}})$ given by (\ref{eq:projRes2}) for every $q_{\boldsymbol{X}}\in\mathcal{E}_{\mathcal{A},\mathcal{B}}^{\leftrightarrow,\perp}$ we can reorganize the divergence as
\begin{eqnarray*}
D(p_{\boldsymbol{X}}|| q_{\boldsymbol{X}}) = D(p_{\boldsymbol{X}} || \stackrel{\rightarrow}{\Pi}_{\mathcal{E}_{\mathcal{A},\mathcal{B}}^{\leftrightarrow,\perp}}(p_{\boldsymbol{X}}))  \\
+ D(\stackrel{\rightarrow}{\Pi}_{\mathcal{E}_{\mathcal{A},\mathcal{B}}^{\leftrightarrow,\perp}}(p_{\boldsymbol{X}}) ||  q_{\boldsymbol{X}_{\mathcal{A}\setminus \mathcal{B}} | \boldsymbol{X}_{\mathcal{A}\cap \mathcal{B}}} q_{\boldsymbol{X}_{\mathcal{B}}} q_{\boldsymbol{X}_{(\mathcal{A}\cup\mathcal{B})^c}})
\end{eqnarray*}
The remainder of the corollary is proved by substituting (\ref{eq:projRes1}) and (\ref{eq:projRes2}) in the equation for the divergence.
\hspace*{\fill} $\Box$

In order to further illustrate these ideas, we will demonstrate Theorem \ref{ig_lemma1}, Theorem \ref{ig_lemma2} and Corollary \ref{thm:proj} in the context of $\overline{\Gamma}^*_3$. 
\begin{example} \label{MainIG_example}Setting $N=3$ we have $\boldsymbol{\mathcal{X}} = \mathcal{X}_1 \times \mathcal{X}_2 \times \mathcal{X}_3$.
Let $\mathcal{N} = \{1,2,3\}$, and denote $\mathcal{A}$ and  $\mathcal{B}$ as some subset of $\mathcal{N}$ that is not equal to $\emptyset$ and $\mathcal{N}$, then we consider the following submanifolds of $\mathcal{O}(\boldsymbol{\mathcal{X}})$:
\begin{eqnarray*}
&& B_{\mathcal{A},\mathcal{B}} = \left\{  p_{\boldsymbol{X}}(\cdot) \left|\ p_{\boldsymbol{X}} = p_{\boldsymbol{X}_{\mathcal{A}}} \cdot p_{\boldsymbol{X}_{\mathcal{B} \backslash \mathcal{A} | (\mathcal{A} \cap \mathcal{B})}}  \right.   \right\} \\
&& Q_{\mathcal{A},\mathcal{B}} = \left\{  p_{\boldsymbol{X}}(\cdot) \left|\ p_{\boldsymbol{X}}= p_{\boldsymbol{X}_{\mathcal{A}}} \cdot p_{\boldsymbol{X}_{\mathcal{B}}} \right.   \right\} \\
&& E_3 =\left\{  p_{\boldsymbol{X}}(\cdot) \left|\ p_{\boldsymbol{X}} = \prod_{i = 1}^{3} p_{X_i} \right.   \right\} \\
&& U_{\mathcal{A}} = \left\{  p_{\boldsymbol{X}}(\cdot) \left|\ p_{\boldsymbol{X}} = \frac{1}{\prod_{i \in \mathcal{A}}|\chi_i|}p_{\boldsymbol{X}_{\mathcal{A}^c}} \right.   \right\} \\
\end{eqnarray*}
Theorem \ref{ig_lemma1} and Theorem \ref{ig_lemma2} show that  $B_{\mathcal{A},\mathcal{B}}$, $Q_{\mathcal{A},\mathcal{B}} $,  $E_3$ and $U_{\mathcal{A}}$ are all nested e-autoparallel submanifolds of $\mathcal{O}(\boldsymbol{\mathcal{X}})$.  For example, $U_{123} \subset U_{23} \subset E_3 \cap U_3 \subset E_3 \subset Q_{13,2} \subset B_{12,13} \subset S_3$, and in this chain, the previous one is the e-autoparallel submanifold of the later one. If we denote $p_{B_{12,13}}$, $p_{Q_{13,2}}$, $p_{E_3}$, $p_{E_3 \cap U_3}$, $p_{U_{23}}$ and $p_{U_{123}}$ the m-projection of a point $p \in S_3 \backslash B_{12,13}$ onto $B_{12,13}$, $Q_{13,2}$, $E_3$, $E_3 \cap U_3$, $U_{23}$ and $U_{123}$ respectively, we will have the Pythagorean relation as shown in Figure \ref{ig_pyth3}.
\begin{figure}
\begin{center}
\includegraphics[width=3in]{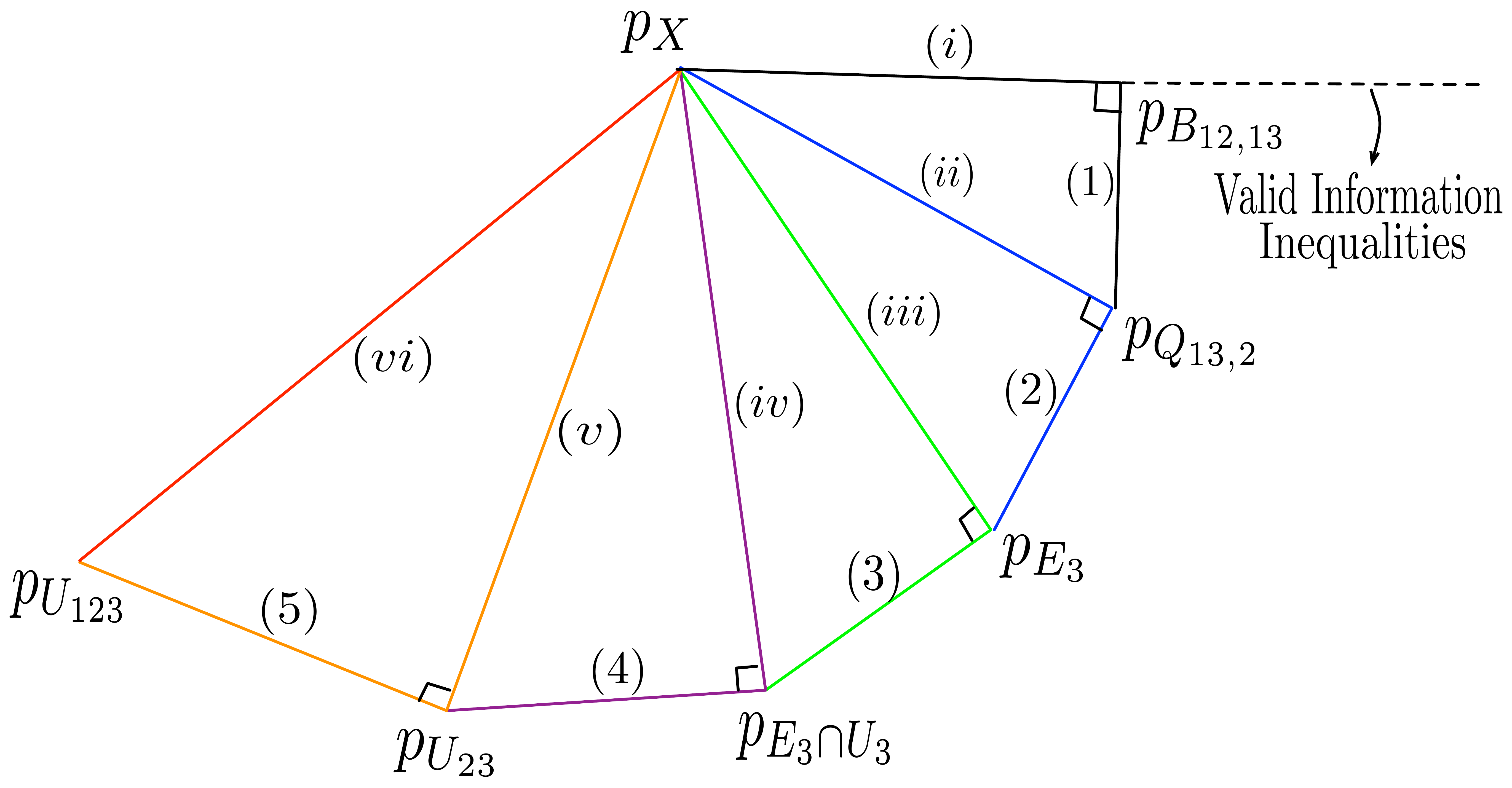}
\caption{Pythagorean relationship on 3 discrete random variables}
\label{ig_pyth3}
\end{center}
\end{figure}
From these pythagorean relations, we will get a series of Information inequalities
\begin{eqnarray*}
&&(\romannumeral 1) \quad h_{12} + h_{13} \geqslant h_{1} + h_{123}\\
&&(\romannumeral 2) \quad h_{13} + h_2 \geqslant h_{123}\\
&&(\romannumeral 3) \quad h_1 + h_2 + h_3 \geqslant h_{123}\\
&&(\romannumeral 4) \quad h_1 + h_2 + \log |\chi_3| \geqslant h_{123}\\
&&(\romannumeral 5) \quad h_1 + \log |\chi_2| + \log |\chi_3| \geqslant h_{123}\\
&&(\romannumeral 6) \quad \log |\chi| \geqslant h_{123}
\end{eqnarray*}
as well as
\begin{eqnarray*}
&&(1) \quad h_{1} + h_{2} \geqslant h_{12}\\
&&(2) \quad h_{1} + h_{3} \geqslant h_{13}\\
&&(3) \quad \log |\chi_3| \geqslant h_{3}\\
&&(4) \quad \log |\chi_2| \geqslant h_{2}\\
&&(5) \quad \log |\chi_1| \geqslant h_{1}
\end{eqnarray*}
where each inequality corresponds to the positivity of the divergence with the associated label in Figure \ref{ig_pyth3}.  As these are
the basic Shannon type inequalities for $N=3$, this diagram yields an information geometric interpretation for the region of entropic vectors $\overline{\Gamma}_3^*$.
\end{example}

Lemma \ref{CIThetaSpace} can also be used to derive information geometric results for certain Shannon faces. 
\begin{theorem} \label{ig_thmface}
Let relation $\mathcal{L}' \subseteq \mathcal{S}(N)$ be the containment minimal $p$-$representable$ semimatroid containing a set $\mathcal{L} \subseteq \mathcal{S}(N)$ such that $\{i \cup j \cup \mathcal{K}\}=\mathcal{N^{\prime}} \subseteq \mathcal{N}$ for $\forall (i,j|\mathcal{K}) \in \mathcal{L}$. Denote $\mathcal{O}(\boldsymbol{\mathcal{X}}^\prime)$ the manifold of probability distributions for random vector $\boldsymbol{X}_{\mathcal{N}^\prime}$.
Let $N^{\prime} = |\mathcal{N}^{\prime}|$ and define $\mathcal{F}_{\mathcal{N}^{\prime}}$ in (\ref{prep_face_constraint_subset}), 
\begin{equation} 
\label{prep_face_constraint_subset}
\mathcal{F}_{\mathcal{N}^{\prime}}= \{ \ \boldsymbol{h} \in \Gamma_{N^{\prime}} \ | \ h_{i\mathcal{K}} + h_{j\mathcal{K}} - h_\mathcal{K} -h_{ij\mathcal{K}} = 0,\ \forall (i,j|\mathcal{K}) \in \mathcal{L}' \}
\end{equation}
then the set of distributions corresponding to $\mathcal{F}_{\mathcal{N}^{\prime}}$ form a e-autoparallel submanifold in $\mathcal{O}(\boldsymbol{\mathcal{X}}^\prime)$. 
\end{theorem}
\noindent \textbf{Proof}: From Corollary \ref{thm:interiorpoints} we know $\mathcal{F}_{\mathcal{N}^{\prime}}$ is a face of $\Gamma_{N^{\prime}}$.
As a result of Lemma \ref{CIThetaSpace}, each $(i,j|\mathcal{K})\in\mathcal{L}$ correspond to a series of hyperplanes (\ref{thetamulti}) in $\boldsymbol{\theta}$ coordinate, we use $E_{eq}^{\mathcal{F}_0}$ to denote the intersection of all the hyperplanes representing any $(i,j|\mathcal{K})$ pairs in $\mathcal{L}$. Since for the submanifold 
\begin{equation*}
\label{Eprep_manifold}
\mathcal{E}^{\perp} = \left\{p_{\boldsymbol{X}}(\cdot) \left|\ p_{\boldsymbol{X}}(\boldsymbol{x})= \prod_{i \in \mathcal{N}^{\prime}} p_{X_i}(x_i) \right. \right\}
\end{equation*}
we have $\boldsymbol{h}(\mathcal{E}^{\perp}) \subseteq \mathcal{F}_{\mathcal{N}^{\prime}}$, which means the intersection $E_{eq}^{\mathcal{F}_0}$ will always be non-empty, and hence there will always exist a $p$-representable semimatroid containing $\mathcal{L}$. The containment minimal $p$-representable matroid will contain any other conditional independence relations implied by those in $\mathcal{L}$, and hence any point in $E_{eq}^{\mathcal{F}_0}$ will obey these too.  Furthermore, $E_{eq}^{\mathcal{F}_0}$ is e-autoparallel in $\mathcal{O}(\boldsymbol{\mathcal{X}}^\prime)$ since it is the non-empty interactions of hyperplanes in $\boldsymbol{\theta}$ coordinate according to Lemma \ref{CIThetaSpace}.
\hspace*{\fill} $\Box$

\begin{figure}
\centering
\includegraphics[width=.47\textwidth]{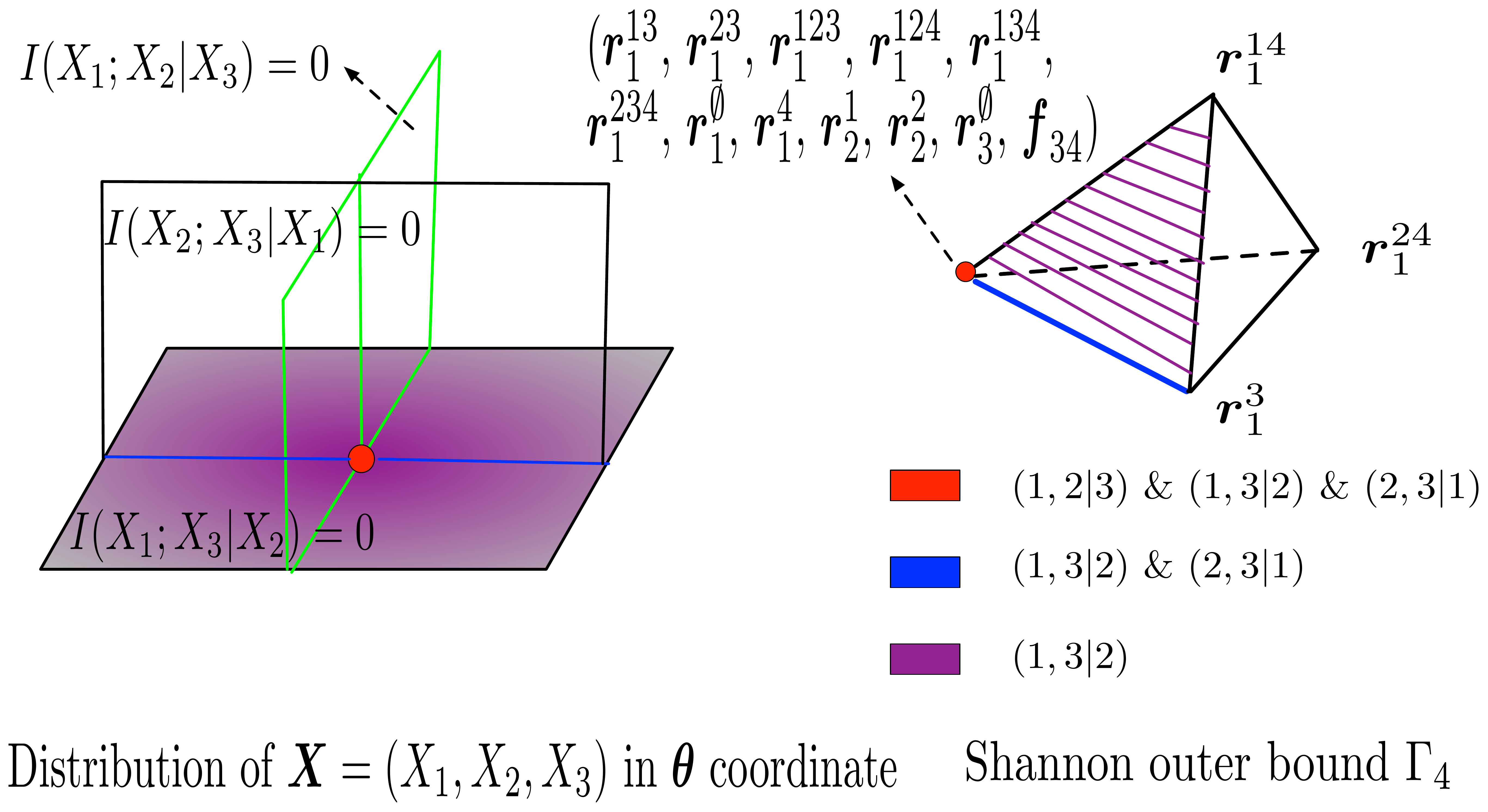}
\caption{The mapping between some faces of $\Gamma_4$ and submanifold of four variable distribution in $\boldsymbol{\theta}$ coordinate}\label{fig:distmapping}
\end{figure}

\begin{example}\label{Example_face}As a example, recall in the last paragraph of section \S \ref{ssec:gap} we study the relationship between subset of the extreme rays of $G^{34}_{4}$ and $p$-$representable$ semimatroids $\{(1,3|2)\}$, $\{(1,3|2) \ \& \ (2,3|1)\}$ and $\{(1,2|3) \ \& \ (1,3|2) \ \& \ (2,3|1)\}$. Since all these three semimatroids are $p$-$representable$, there are distributions satisfy the conditional independent relations corresponding to these $(i,j | \mathcal{K})$ couples. For example, the distributions corresponding to semimatroid $\{(1,2|3) \ \& \ (1,3|2) \ \& \ (2,3|1)\}$ satisfied the constraints $I (X_1;X_2|X_3)= 0$,  $I (X_1;X_3|X_2) = 0$ and $I (X_2;X_3|X_1) = 0$. 
Then according to Lemma \ref{CIThetaSpace}, the distributions which corresponding to semimatroid $\{(1,2|3) \ \& \ (1,3|2) \ \& \ (2,3|1)\}$ must obey (\ref{thetamulti}) defined for each of the three $(i,j|\mathcal{K})$ couples. In $\boldsymbol{\theta}$ coordinate, the mapping can be illustrated by the plot on the left of Fig. \ref{fig:distmapping}, in which we only consider three random variables $X_1$, $X_2$ and $X_3$. Since each of three set of distributions is e-autoparallel, any their intersections are also e-autoparallel.
On the right of Fig. \ref{fig:distmapping} we plot the relationship among the Shannon faces corresponding to the three $p$-$representable$ semimatroids $\{(1,3|2)\}$, $\{(1,3|2) \ \& \ (2,3|1)\}$ and $\{(1,2|3) \ \& \ (1,3|2) \ \& \ (2,3|1)\}$ in the gap $G^{34}_{4}$ of $\Gamma_4$. As shown in Fig. \ref{fig:distmapping}, the polyhedral cone constructed from $\boldsymbol{r}_1^{3}$, $\boldsymbol{r}_1^{14}$, $\boldsymbol{r}_1^{13}$, $\boldsymbol{r}_1^{23}$,  $\boldsymbol{r}_1^{123}$, $\boldsymbol{r}_1^{124}$, $\boldsymbol{r}_1^{134}$, $\boldsymbol{r}_1^{234}$, $\boldsymbol{r}_1^{\emptyset}$, $\boldsymbol{r}_1^{4}$, $\boldsymbol{r}_2^{1}$, $\boldsymbol{r}_2^{2}$, $\boldsymbol{r}_3^{\emptyset}$ and $\boldsymbol{f}_{34}$ corresponding to semimatroids $\{(1,3|2)\}$; 
removing $\boldsymbol{r}_1^{14}$ from the list, we get semimatroids $\{(1,3|2) \ \& \ (2,3|1)\}$; 
finally, the intersection of three faces corresponding to the polyhedral cone constructed from $\boldsymbol{r}_1^{13}$, $\boldsymbol{r}_1^{23}$,  $\boldsymbol{r}_1^{123}$, $\boldsymbol{r}_1^{124}$, $\boldsymbol{r}_1^{134}$, $\boldsymbol{r}_1^{234}$, $\boldsymbol{r}_1^{\emptyset}$, $\boldsymbol{r}_1^{4}$, $\boldsymbol{r}_2^{1}$, $\boldsymbol{r}_2^{2}$, $\boldsymbol{r}_3^{\emptyset}$ and $\boldsymbol{f}_{34}$.
\end{example}
As for the relations between $p$-$representable$ semimatroids and the conditions of $\mathcal{L}$ specified in Theorem \ref{ig_thmface}, by examine the 120 irreducible $p$-$representable$ semimatroids listed in Theorem \ref{theoremmatus}, we get the following Corollary:
\begin{corollary} \label{thm:ingletonface}
Let relation $\mathcal{L} \subseteq \mathcal{S}(4)$ be a $p$-$representable$ semimatroid such that $\mathcal{L}$ contains at least two $(i,j|\mathcal{K})$ pairs. If $\{i \cup j \cup \mathcal{K}\}=\mathcal{N^{\prime}} \subseteq \{1,2,3,4\}$ for $\forall (i,j|\mathcal{K}) \in \mathcal{L}$, then $\mathcal{L}$ is an Ingleton semimatroid.
\end{corollary}
In this sense, the information geometric parametrization presented thus far can only reach those faces of $\bar{\Gamma}^*_4$ shared with the Ingleton inner bound.  In part for this reason, in the next section we will study the information geometric structure of certain Ingleton violating distributions.

\vspace{-4mm}
\subsection{Information Geometric Structure of Ingleton-Violating Entropic Vectors \& their Distributions}\label{ssec:IngVio}


As discussed in \S \ref{ssec:ingleVio}, every $k$-atom distribution for $k<4$ is incapable of violating Ingleton, and every $k$-atom distribution for $k\in\{5,6,7\}$ are incapable of exceeding the Ingleton score achieved by the $4$-atom distribution on support $\boldsymbol{\mathcal{X}}_{4}$ given in (\ref{eq:4atoms}).  Here we investigate study the information geometric structure of those distributions violating Ingleton on this special $4$-atom support, and show that they admit a nice information geometric.  We will also show that larger $k$-atom supports that are capable of violating Ingleton, but do not yield better Ingleton scores, do not share this special information geometric structure.

\begin{figure}
\centering
\includegraphics[width=.36\textwidth]{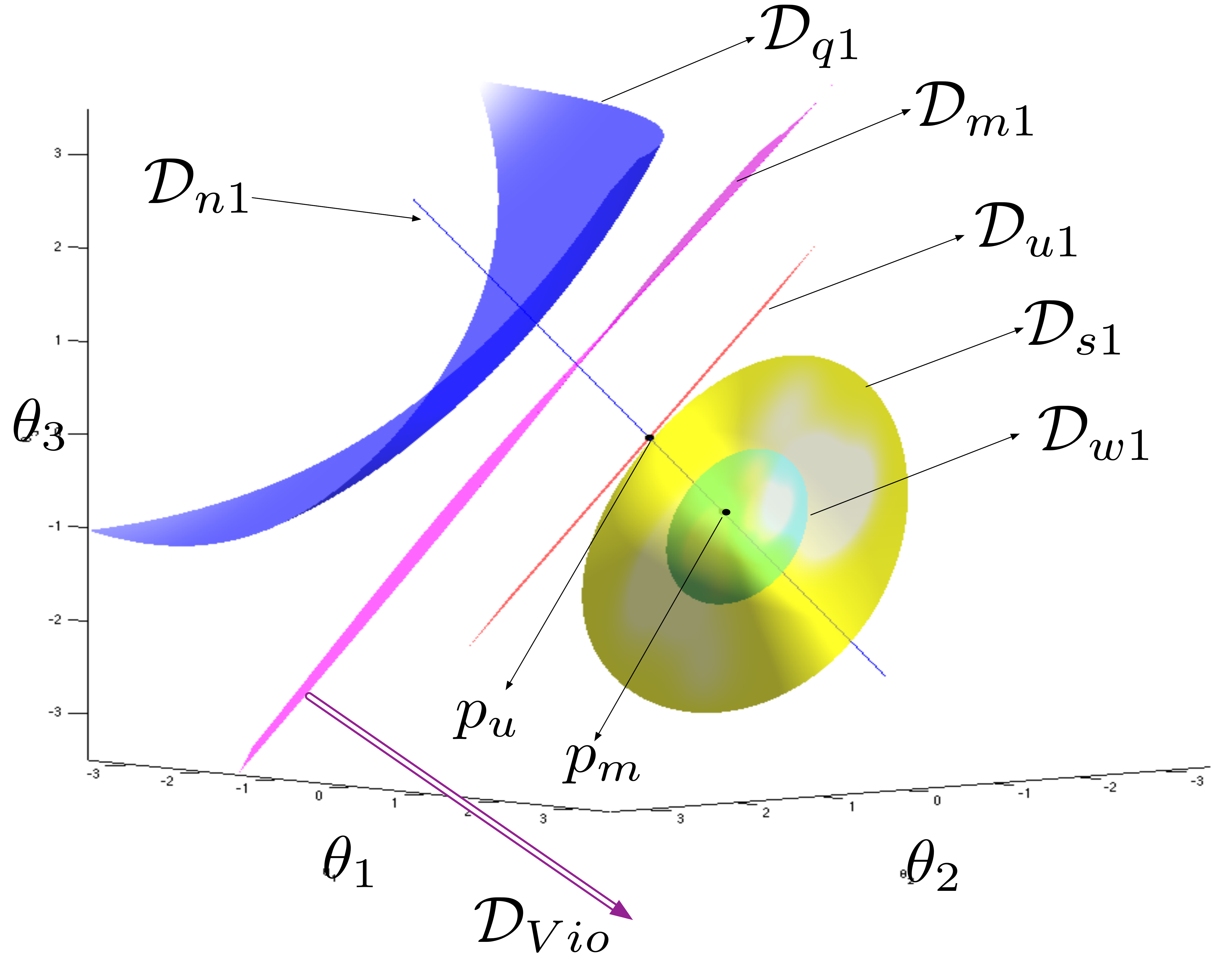}
\caption{Manifold $\mathcal{D}_{4atom}$ in $\theta$ coordinate}\label{fig:thetaplot}
\end{figure}

Denote by $\mathcal{O}(\boldsymbol{\mathcal{X}}_4) $ the manifold of all probability mass functions for four binary random variables with the support  (\ref{eq:4atoms}).  Parameterize these distributions with the parameters $\eta_1 = \mathbb{P}(\boldsymbol{X} = 0000) = \alpha$, $\eta_2 = \mathbb{P}(\boldsymbol{X} = 0110) = \beta - \alpha$, $\eta_3= \mathbb{P}(\boldsymbol{X} = 1010) = \gamma - \alpha$,  $\eta_4 = \mathbb{P}(\boldsymbol{X} = 1111) = 1 - \sum^{3}_{i = 1} \eta_i$ = $1 + \alpha - \beta - \gamma$, yielding the m-coordinates of $\mathcal{O}(\boldsymbol{\mathcal{X}}_4)$. Let the associated $e$-coordinates can be calculated as $\theta_i = \log_{2} \frac{\boldsymbol{\eta}_i}{\boldsymbol{\eta}_4}$ for $i = 1,2,3$. Next, we consider the submanifold $\mathcal{D}_{u1}$ = $\{p_x \in \mathcal{O}(\boldsymbol{\mathcal{X}}_4) | I(X_3;X_4) = 0\}$, by Theorem \ref{ig_lemma1}, $\mathcal{O}(\boldsymbol{\mathcal{X}}_4)$ is a e-autoparallel submanifold of $\mathcal{D}_{4atom}$.  In fact, an equivalent definition is $\mathcal{D}_{u1} = \{ p_x \in \mathcal{O}(\boldsymbol{\mathcal{X}}_4) | -\theta_1 + \theta_2 + \theta_3 = 0 \}$. Numerical calculation illustrated in the Figure \ref{fig:thetaplot} lead to the following proposition, which we have verified numerically.
\vspace{-2mm}
\begin{proposition} \label{ig_lemma3}
Let $\mathcal{D}_{m1}$ = $\{p_x \in \mathcal{O}(\boldsymbol{\mathcal{X}}_4) | Ingleton_{34}= 0\}$, then $\mathcal{D}_{m1}$ is a e-autoparallel submanifold of $\mathcal{O}(\boldsymbol{\mathcal{X}}_4)$ and is parallel with $\mathcal{D}_{u1}$ in $e$-coordinate.  In fact, an equivalent definition of $\mathcal{D}_{m1}$ is $\mathcal{D}_{m1}$ = $\{ p_x \in \mathcal{O}(\boldsymbol{\mathcal{X}}_4) | -\theta_1 + \theta_2 + \theta_3 = \log_{2} (\frac{0.5 - \alpha_{0}}{\alpha_{0}})^{2} \}$, where $\alpha_{0}$ is the solution of $- \alpha \log_{2}\alpha - (1- \alpha)\log_{2}(1- \alpha) = \frac{1+2 \alpha}{2}$ in $0 < \alpha < \frac{1}{2}$.
\end{proposition}
\vspace{-2mm}
In fact, using this equivalent definition, we can also determine all the distributions in $\mathcal{O}(\boldsymbol{\mathcal{X}}_4)$ that violate $Ingleton_{34} \geqslant 0$ as the submanifold $\mathcal{D}_{Vio}$ = $\{p_x \in \mathcal{O}(\boldsymbol{\mathcal{X}}_4) | -\theta_1 + \theta_2 + \theta_3 < \log_{2} (\frac{0.5 - \alpha_{0}}{\alpha_{0}})^{2} \}$. Because we are dealing with $\mathcal{O}(\boldsymbol{\mathcal{X}}_4) $, a 3 dimensional manifold, we can use a plot to visualize our results in Fig. \ref{fig:thetaplot}. In Fig.  \ref{fig:thetaplot}, besides $\mathcal{D}_{u1}$ and $\mathcal{D}_{m1}$, we also plot the following submanifolds and points:
\begin{eqnarray*}
\mathcal{D}_{q1} &&= \{p_x \in \mathcal{O}(\boldsymbol{\mathcal{X}}_4) \ | \ Ingleton_{34} = 0.1\} \\
\mathcal{D}_{s1} &&= \{p_x \in \mathcal{O}(\boldsymbol{\mathcal{X}}_4)\ | \ Ingleton_{34} = -0.126\} \\
\mathcal{D}_{w1} &&= \{p_x \in \mathcal{O}(\boldsymbol{\mathcal{X}}_4) \ | \ Ingleton_{34}= -0.16\} \\
\mathcal{D}_{n1} &&= \{p_x \in \mathcal{O}(\boldsymbol{\mathcal{X}}_4) \ | \ \beta = \gamma = 0.5 \} \\
p_{u} &&=  \{p_x \in \mathcal{O}(\boldsymbol{\mathcal{X}}_4) \ | \ \alpha = 0.25, \beta = \gamma = 0.5 \} \\
p_{m} &&=  \{p_x \in \mathcal{O}(\boldsymbol{\mathcal{X}}_4) \ | \ \alpha \approx 0.33, \beta = \gamma = 0.5 \}
\end{eqnarray*}
As we can see from Fig.  \ref{fig:thetaplot}, $\mathcal{D}_{m1}$ and $\mathcal{D}_{u1}$ are e-autoparallel and parallel to each other. As $Ingleton_{34}$ goes from 0 to negative values, the hyperplane becomes ellipsoid-like, and as $Ingleton_{34}$ becomes smaller and smaller, the ellipsoid-like surface shrinks, finally shrinking to a single point $p_{m}$ at $Ingleton_{34} \approx -0.1699$, the point associated with the four atom conjecture in \cite{DFZ_IneqsPaper}. Also for each e-autoparallel submanifold $\mathcal{D}_{\forall e}$ $\subset$ $\mathcal{D}_{Vio}$
that is parallel to $\mathcal{D}_{m1}$ in $e$-coordinate, the minimum Ingleton of $\mathcal{D}_{\forall e}$ is achieved at point $\mathcal{D}_{\forall e} \cap \mathcal{D}_{n1}$, where $\mathcal{D}_{n1}$ is the $e$-\emph{geodesic} in which marginal distribution of $X_3$ and $X_4$ are uniform, i.e. $\beta = \gamma = 0.5$.

\begin{figure}
\centering
\includegraphics[width=.22\textwidth]{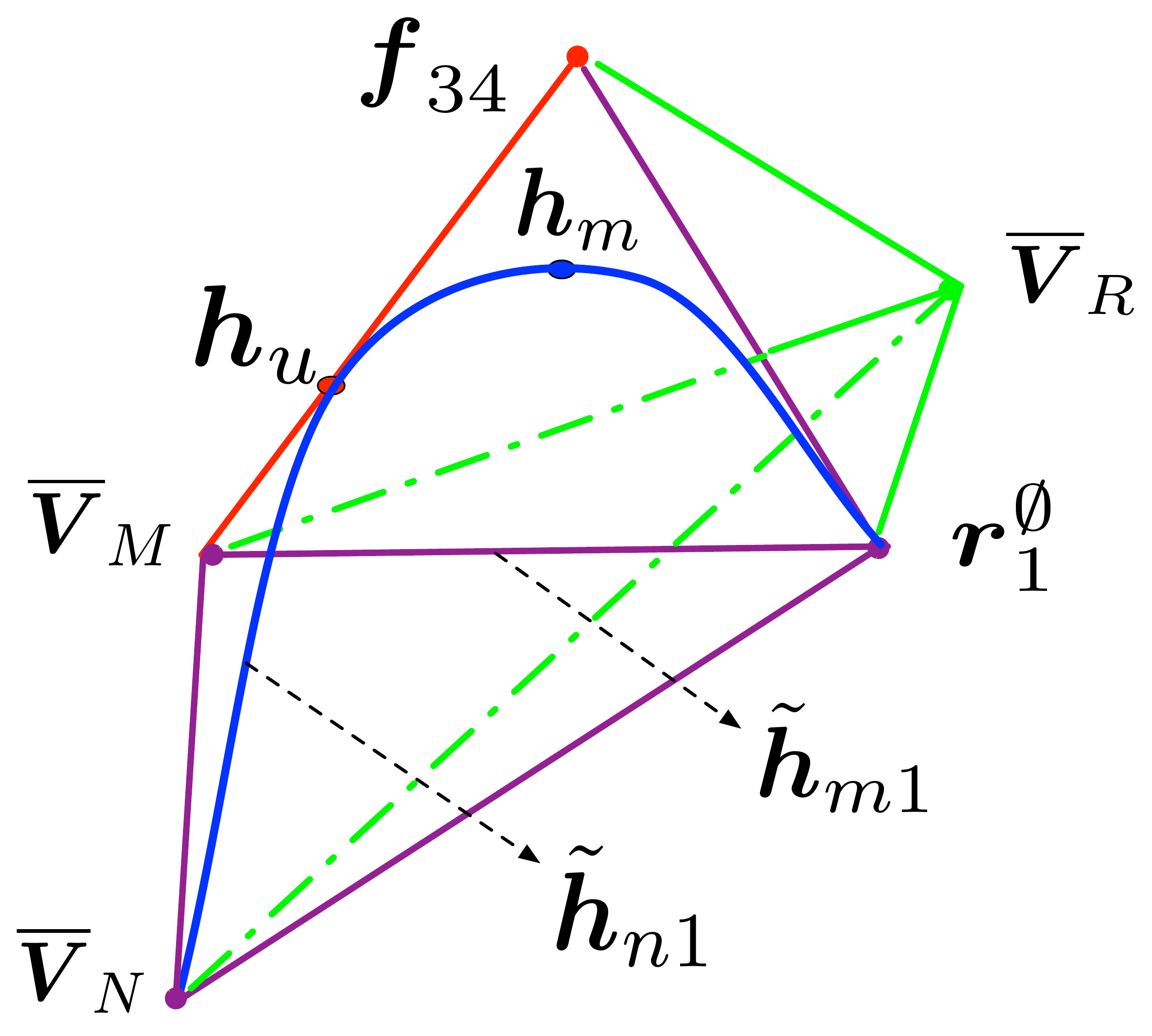}
\caption{$G^{34}_{4}$: one of the six gaps between $\mathcal{I}_4$ and $\Gamma_4$}\label{fig:gap}
\end{figure}

\vspace{1mm}
Next, we map some submanifolds of $\mathcal{O}(\boldsymbol{\mathcal{X}}_4)$ to the entropic region. From Lemma \ref{lemmamatus}, $G^{34}_{4}$, one of the six gaps between $\mathcal{I}_4$ and $\Gamma_4$ is characterized by extreme rays $\overline{\boldsymbol{V}}_{P} = (\overline{\boldsymbol{V}}_M,\overline{\boldsymbol{V}}_R,\boldsymbol{r}^{\emptyset}_1, \boldsymbol{f}_{34})$,
where $\overline{\boldsymbol{V}}_M = (\boldsymbol{r}^{13}_1,\boldsymbol{r}^{14}_1,\boldsymbol{r}^{23}_1,\boldsymbol{r}^{24}_1,\boldsymbol{r}^{1}_2,\boldsymbol{r}^{2}_2)$ and $\overline{\boldsymbol{V}}_R = (\boldsymbol{r}^{123}_1,\boldsymbol{r}^{124}_1,\boldsymbol{r}^{134}_1,\boldsymbol{r}^{234}_1,\boldsymbol{r}^{3}_1,\boldsymbol{r}^{4}_1,\boldsymbol{r}^{\emptyset}_3)$. In addition, we define $\overline{\boldsymbol{V}}_N = (\boldsymbol{r}^{1}_1,\boldsymbol{r}^{2}_1,\boldsymbol{r}^{12}_1)$, and use Fig. \ref{fig:gap} to help visualize $G^{34}_{4}$.

\vspace{1mm}
\indent In Fig. \ref{fig:gap}, $\boldsymbol{f}_{34}$ is one of the six Ingleton-violating extreme ray of $\Gamma_4$, $\overline{\boldsymbol{V}}_M$,$\overline{\boldsymbol{V}}_R$ and $\boldsymbol{r}^{\emptyset}_1$ are all extreme rays of $\mathcal{I}_4$ that make $Ingleton_{34} = 0$. Based on the information geometric characterization, the mapping from $\mathcal{O}(\boldsymbol{\mathcal{X}}_4)$ to $\Gamma^{*}_4$ is straight forward: the curve $\tilde{\boldsymbol{h}}_{n1}$ = $\boldsymbol{h}(\mathcal{D}_{n1})$, the straight line $\tilde{\boldsymbol{h}}_{m1}$ = $\boldsymbol{h}(\mathcal{D}_{m1})$, the point $\boldsymbol{h}_{u}$ = $\boldsymbol{h}(p_{u})$ and the point $\boldsymbol{h}_{m}$ = $\boldsymbol{h}(p_{m})$. \\

\begin{figure}
\centering
\includegraphics[width=.43\textwidth]{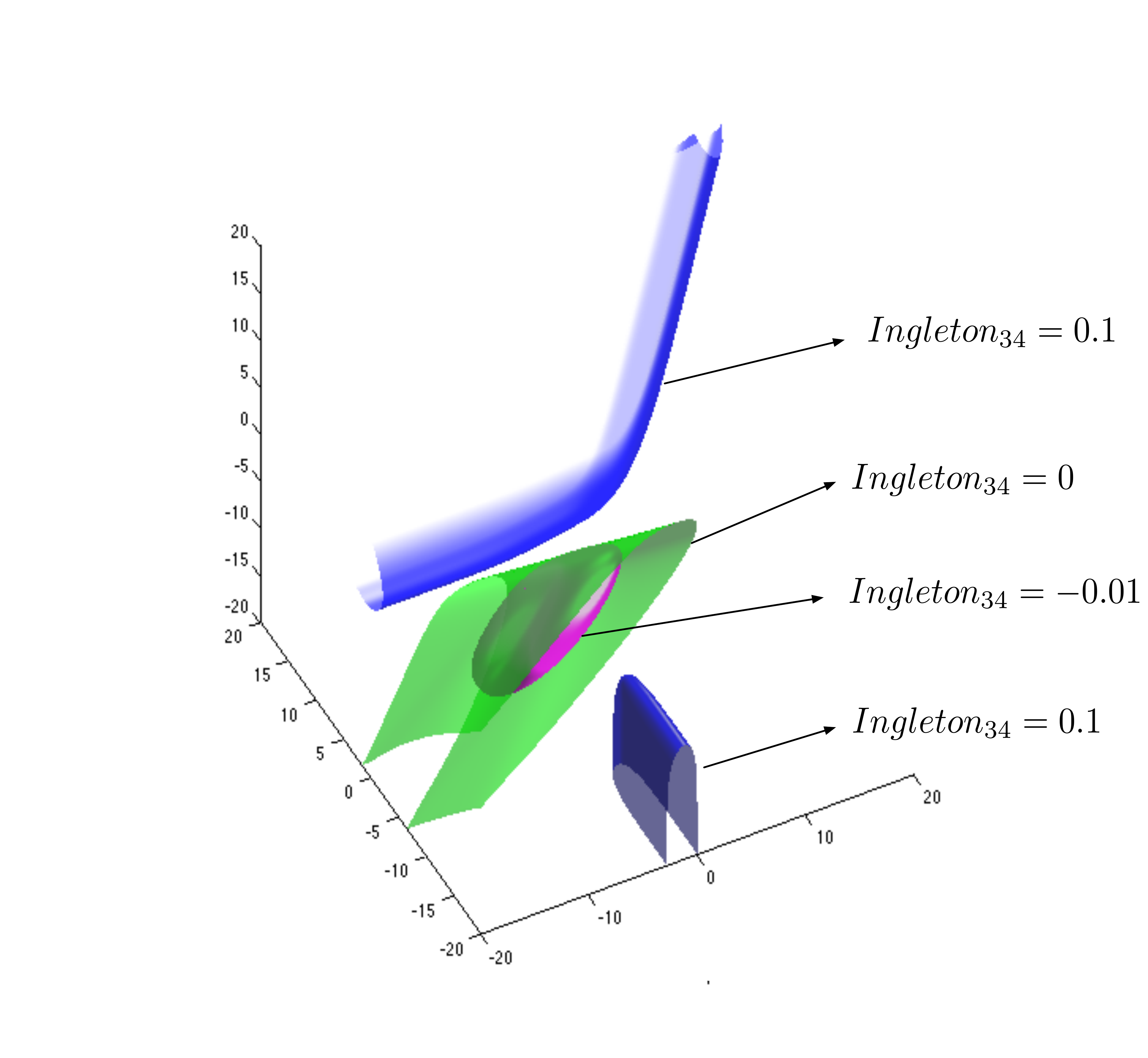}
\caption{Slice of $\mathcal{D}_{5atom}$ in $\theta$ coordinate}\label{fig:5atomIngleton}
\end{figure}

Given this nice information geometric property of $4$-atom support that Ingleton violation corresponds to a half-space, a natural question to ask is if the e-autoparallel property of $\mathcal{D}_{m1}$ can be extended to five and more atoms. Since we already obtain the list of 29 nonisomorphic $5$-atom distribution supports, we know among the 29 supports, only one of them as defined in (\ref{eq:5atoms}) is not a direct extension of $4$-atom distributions. For this $5$-atom support, if we fix one coordinate $\theta_{0}$ in $e$-coordinate, the resulting manifold will be three dimension, thus we can plot the hyperplane of $Ingleton_{34}$ = 0 to check if it is e-autoparallel. The result is shown in Figure \ref{fig:5atomIngleton}. As we can see, the curvature of $Ingleton_{34} = 0$ is non-zero, thus it can not be e-autoparallel. However, the property that distributions of $Ingleton_{34}> 0$ on one side of $Ingleton_{34} = 0$, and distributions of $Ingleton_{34}< 0$ on the other side of $Ingleton_{34} = 0$ still hold.

\vspace{1mm}

\vspace{-2.2mm}
\section{Conclusions}\label{sec:conculsions}
In this paper, we reviewed the importance of the region of entropic vectors in applications ranging from multiterminal data compression to network coding to multimedia transmission.  The best known bounds for this set, which primarily are based on constructions involving representable matroids and hence linear constructions, were reviewed.  
In order to provide an exhaustive search of more complicated constructions, we proposed and solved the problem of listing non-isomorphic distribution supports for the purpose of calculating entropic vectors. This is carried out by fixing $k$, the number of atoms and $N$, the number of random variables, so we can grow in $k$ or $N$ to see the progress we make towards the characterization of entropy region. Along the way, a recursive algorithm, Snakes and Ladders, was used to efficiently enumerate the unique supports. The concept of inner bounds based on $k$-atom distributions was introduced to aid understanding the structure of the entropic vector region. We experimentally generated $k$-atom inner bounds for $k$ = 4, 5, and 6,  calculated the volume of these inner bounds, and visualized them via a certain three dimensional projection. A future research direction in this line of work is to study the conditional independence relations specified by the $k$-atom support that violate Ingleton, and to explore the algebraic structure of $k$-atom supports.

The second part of the paper shifted away from numerical distribution searches toward analytical characterization of properties of distributions which enabled them to be extremal in the sense achieving entropic vectors of living in certain faces of the Shannon outer bound, as well as for violating Ingleton.  These analytical characterizations made use of Information geometric parameterizations.  It was shown that the set of distributions on the support associated with the yet-best Ingleton score achievable with an (not-almost) entropic vector which violate Ingleton correspond to a half-space in an appropriate coordinate system.  This property was shown not be be shared by a larger $5$-atom support achieving a strictly poorer Ingleton score.

\vspace{-4mm}
\bibliographystyle{IEEEtran}
\bibliography{IEEEabrv,myPubsWLinks,entFunc}

\end{document}